\documentclass[aip,
amsmath,amssymb,%
reprint,
twocolumn,
a4paper
]{revtex4-1}

\usepackage[utf8]{inputenc}
\usepackage[T1]{fontenc}

\usepackage{siunitx, physics, bm, mathptmx}

\usepackage{etoolbox, graphicx, xcolor, hyperref, multirow, booktabs, enumerate}
\usepackage[centering,hmargin=2cm,vmargin=2.5cm]{geometry}
\usepackage[caption=false]{subfig}
\usepackage[super]{nth}

\hypersetup{
    colorlinks,
    linkcolor={red!50!black},
    citecolor={blue!50!black},
    urlcolor={blue!80!black}
}

\AtBeginDocument{%
  \heavyrulewidth=.08em
  \lightrulewidth=.05em
  \cmidrulewidth=.03em
  \belowrulesep=.65ex
  \belowbottomsep=0pt
  \aboverulesep=.4ex
  \abovetopsep=0pt
  \cmidrulesep=\doublerulesep
  \cmidrulekern=.5em
  \defaultaddspace=.5em
}

\makeatletter
\def\@email#1#2{%
 \endgroup
 \patchcmd{\titleblock@produce}
  {\frontmatter@RRAPformat}
  {\frontmatter@RRAPformat{\produce@RRAP{*#1\href{mailto:#2}{#2}}}\frontmatter@RRAPformat}
  {}{}
}%
\makeatother

\begin{document}

\title{High-resolution measurements and simulations of the flow through a packed bed and its freeboard region}

\author{Wojciech Sadowski}
\email{wojciech.sadowski@rub.de}
\author{Mohammed Sayyari}
\author{Francesca di Mare}
\affiliation{%
  Chair of Thermal Turbomachines and Aeroengines, 
  Ruhr-Universit\"{a}t Bochum,
  Universit\"{a}tsstra{\ss}e 150 D-44801 Bochum, Germany
}

\author{Christin Velten}
\author{Katharina Z\"{a}hringer}
\affiliation{%
  Lehrstuhl f\"{u}r Str\"{o}mungsmechanik und Str\"{o}mungstechnik,
  Otto-von-Guericke-Universit\"{a}t Magdeburg, 39106 Magdeburg, Germany
}

\date{\today}

\begin{abstract}
    The present study focuses on the assessment of the performance of a Finite Volume Method based,
    particle-resolved simulation approach to predict the flow through a model 
    packed-bed consisting of 21 layers of spheres arranged in
    the body centred cubic packing. The unsteady flow developing in the freeboard is also considered.
    Two highly-resolved large eddy simulation were preformed, for two Reynolds numbers,
    300 and 500, based on the particle diameter, employing a polyhedral, 
    boundary-conforming mesh.
    The geometry and the flow conditions are
    set to reproduce the flow conditions investigated in the experiment carried out
    by \citet{velten2023} using Particle Image Velocimetry. 
    The numerical results compare favourably with the measurements
    both inside and above the bed. The effect of differences arising
    between the physical and numerical configurations are thoroughly discussed 
    alongside the impact of meshing strategy on the accuracy of 
    the predictions.
\end{abstract}

\maketitle

\section{Introduction}

Packed bed reactors represent a class of devices often used in chemical and
process engineering, in which an assembly of particles is passed by a gas,
usually ensuring appropriate mixing of species, introducing reactants or
promoting processes such as drying or coating. The flow characteristics
influence the phenomena occurring in these devices to a great extent. In
particular, the effects of the intermittent coupling between the turbulent flow
in the freeboard, a region above the particle assembly, and the interstitial
flow between the particles has been seen to reach deep inside the packed bed
\citep{shiehnejadhesar2014}.

Numerical simulations can supply extremely valuable insight in the development
of the flow in packed beds to various levels of accuracy, depending on the
methodological approach. On one hand, the porous media approach provides a
simplified, homogenised representation of the particle assembly, whereby the
porosity, generally considered a scalar value, denotes he volume fraction
occupied by the particles \citep{dixon2020} and offers therefore an approximate
description of the interstitial flow. On the other hand, in Particle Resolved
Simulations (PRS), \citep{dixon2006,dixon2020,jurtz2019}, a detailed
description of the passing flow and of the particles is pursued, requiring in
particular appropriate turbulence closure. In the works by
\citealt{shams2013,shams2013a,shams2013b,shams2013c,shams2015} exhaustive
reviews of turbulence modelling strategies in packed beds are provided
alongside a discussion of the nature of turbulence in the interstices by
\citet{jin2015,uth2016} as well as \citet{rao2022}. Direct Numerical Simulation
(DNS), entails the explicit resolution of all relevant length and time scales,
ranging from those smaller the characteristic interstice dimensions and related
to the turbulence length scale in the bed, to the overall dimension of the
reactor, and are associated to prohibitively high computational costs
\citep{lage2002,dixon2020}, above all in technologically relevant applications.
Relaxing the spatial and temporal resolution requirements intrinsic of PRS is
not a straightforward task. For example, an insufficient spatial resolution
near the bed surface could lead to an inaccurately predicted mean velocity in
the whole region occupied by the solid phase and a greatly overestimated drag
force induced by the particles \citep{sadowski2023a}. In \citet{breugem2005}
and \citet{kuwata2016} DNS of flows over periodically repeating sets of
obstacles with, respectively, a Finite Volume Method (FVM) and
Lattice--Boltzmann-based solvers were presented. In the context of the
finite-volume method, \citet{shams2013c,shams2013b} conducted a DNS of a single
cubic pebble bed configuration with heat transfer , revealing highly complex
temperature and velocity distributions under turbulent conditions. Shams et
al.\citealt{shams2014_stacked} performed a Large Eddy Simulation (LES) of the
flow through a randomly stacked pebble bed using FVM and a polyhedral
boundary-conforming mesh. A major issue arises when performing FVM-based PRS
from the necessity of resolving the small gaps between particles with a
reasonable accuracy maintaining an acceptable mesh quality \cite{dixon2020}.
This problem is notoriously difficult to solve, and if untreated, can hinder
the stability and accuracy of the computation \cite{rebughini2016}.

The Immersed Boundary Method (IBM) is often used in conjunction with FVM to
simulate packed bed reactors, as it enables the use of simpler meshes, often of
uniform size or refined by cell-splitting. The particle boundary is represented
as virtually intersecting the fluid cells and its influence on the flow field
is considered by applying a force field in the fluid domain. For example,
\citet{gorges2024comparing} simulated a packed bed using two implementations of
the IBM with smooth and blocked-off particle boundaries. For similar reasons,
the Lattice--Boltzmann Method (LBM) represents another useful alternative as
showed by \citet{neeraj2023a} who simulated the flow inside and over the
surface of a packed bed. Reference experimental data for verification of such
simulations is usually scarce due to the difficulty in accessing the interior
of the packing. Hence, only limited validation of the numerical simulation is
possible using generally global (macroscopic) quantities obtained from
empirical correlations or very sparse and uncertain data. \citet{wongkham2020}
(and references therein) used empirical correlations for the distribution of
porosity near the wall and for the pressure drop along the packed-bed to
validate the implementation of a novel IBM. Also, the experimental measurements
of superficial velocity and porosity by \citet{giese1998} were used by
\citet{wehinger2016} to validate the FVM simulation approach for catalytic flow
reactors. To the best of the authors' knowledge,  studies validating numerical
methods w.r.t.\ the accurate prediction of small-scale interstitial flow and
the phenomena occurring in the freeboard are hardly attempted, mainly due to
the scarcity of experimental data. However, recently data measured by
\citet{velten2023} in an experimental packed-bed configuration were used to
assess both LBM- and IBM-based approaches, which have been shown to produce
satisfactory results when it comes to the prediction of interstitial and
freeboard flow \cite{neeraj2023a,gorges2024comparing}. Although the FVM
approach remains a staple, both in industry and academia, its performance for
PRS has not been assessed to the same extent. Experimental data on these kinds
of flows can be generated by applying intrusive techniques such as hot wire
anemometry \citep{Morales1951,Schwartz1953} or endoscopic Particle Image
Velocimetry (PIV) \citep{Blois2012} to investigate the flow through the
packing. The usual drawback of these methods lies in the fact that the
insertion of the measurement device into the packed-bed disturbs the flow.
Fortunately, other nonintrusive methods are also available. For example,
tomographic techniques such as Magnetic Resonance Imaging (MRI)
\citep{Sederman1997,Suekane2003,Lovreglio2018} or optical techniques such as
planar PIV \citep{Pokrajac2009,Khayamyan2017a,Khayamyan2017b} or tomographic
PIV \citep{Larsson2018} can be applied. While tomographic techniques are mainly
limited to liquid flows and do not need optical accessibility, as it is the
case for optical methods, the resolution in space and time is comparably low
\citep{Gladden2006,Poelma2020}. On the other hand, for optical techniques,
transparent materials have to be used to allow for measurement access. When
employing transparent packing material, however, distortion is introduced
which, in the case of liquids, is usually handled by refractive index matching
\citep{Hassan2008}. Unfortunately, this method is obviously not available for
gaseous flows. Hence, most existing data on flow fields inside packed beds is
based on liquids. Some approaches to overcome the issue in gases exist, either
by additional post-processing, e.g., by calibration \citep{Kovats2015}, or a
ray tracing based correction \citep{Martins2018,Velten2024}. Nevertheless,
standard PIV can also be applied to the gas flow in regions above the packed
bed as well as small regions in the interior of the bed, specially if a regular
arrangement, such as the body-centred cubic (BCC) packing, is investigated
\citep{velten2023}. 

In this paper, we conduct particle-resolved Large Eddy Simulations for two
particle Reynolds numbers in a vertical packed bed configuration passed by a
gas stream using Finite Volume Method. Gas velocity data gathered by PIV at the
outlet and in the optically accessible interstices of a BCC packing
\citep{velten2023} is used for validation and in order to facilitate the
analysis of the computed results. The above mentioned optical obstructions
limit the potential of the PIV data, however, the relatively idealised setup of
a numerical simulation opens the door for extracting the results in regions
inaccessible to current experimental measurement techniques. 
sentence be still here? The analysed geometry is based on the experimental
set-up investigated by \citet{velten2023}. Both the interstitial flow and the
unsteady velocity field above the surface of the bed is analysed. By directly
comparing the predicted flow field to the measurements we demonstrate the
capabilities and shortcomings of the classical boundary conforming FVM approach
for PRS. We also discuss the potential effects of unavoidable discrepancies
between the numerical and real geometrical arrangements on the prediction of
the velocity field (see Section~\ref{sec:math-model}). Finally, we analyse the
computed results from the point of view of the influence of the wall
channelling effects on the flow conditions inside the packing and the
interactions between the flow in the freeboard and velocity inside the bed. The
paper is structured as follows: first, the geometry and flow conditions are
described in Section~\ref{sec:experimental-configuration}, then, in
Section~\ref{sec:math-model}, the numerical setup is thoroughly discussed,
followed by the description of the characteristics of the PIV equipment and
positions of the measurement regions in Section~\ref{sec:piv-setup}. Finally,
the results inside the packing and over the freeboard are described and
discussed in Section~\ref{sec:results}. Concluding remarks and an outlook of
future research are summarised in Section~\ref{sec:conclusions}.

\section{Geometry and flow conditions}
\label{sec:experimental-configuration}
\begin{figure}[t]
\centering
\subfloat[]{%
    \label{fig:reactor_geometry}%
    \includegraphics[height=10cm]{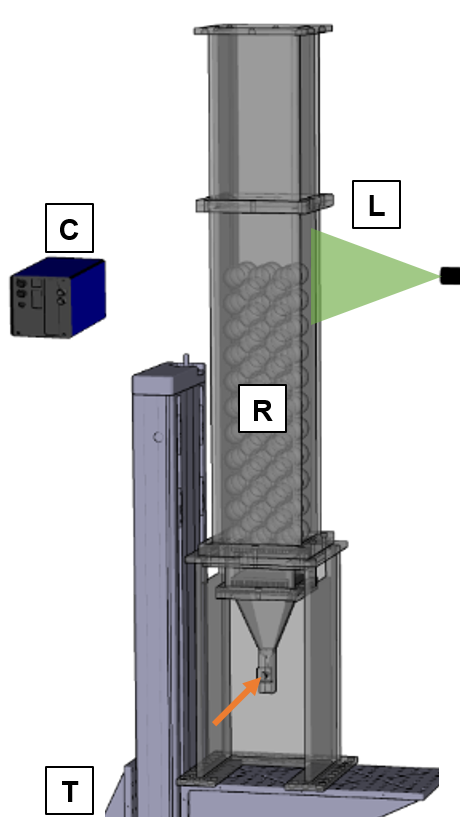}
}\\
\subfloat[]{%
    \label{fig:base_plate}%
    \includegraphics[height=4cm]{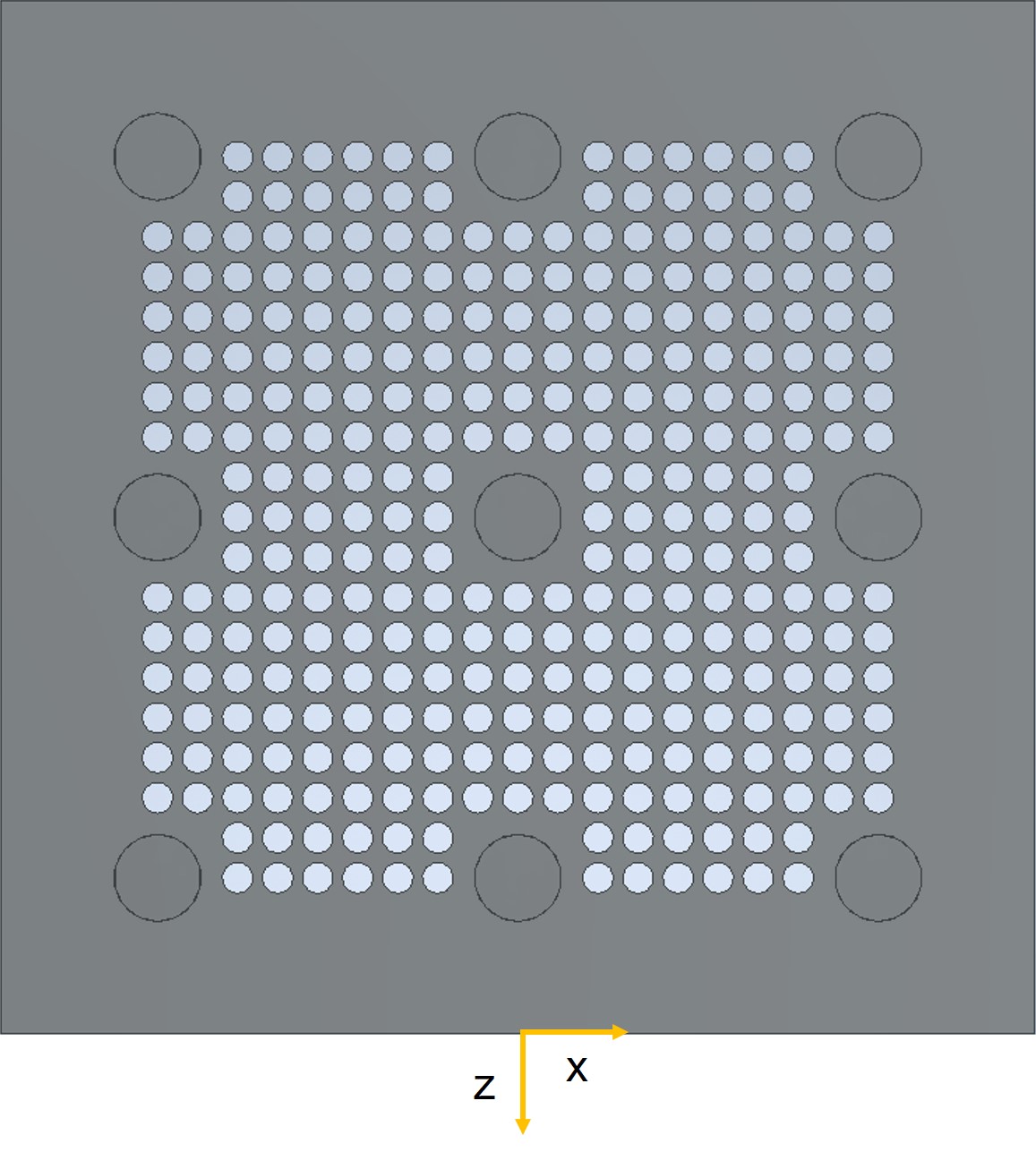}
}
\caption{Schematic of the geometry of the experimental configuration. In the
  top figure: the experimental test section (R) with the indicated position of the light
  sheet (L), the camera (C), and the traversing system (T). The packing houses
  21 layers of spheres in a BCC structure. The lower picture presents the
  test section's baseplate with gas inlet holes (white circles) and bearings (black
  circles) housing the first layer of spheres. The axes shown in the picture
  represent the coordinate system of the PIV data, with the \(y=0\) plane
  coincident with the top surface of the baseplate and the \(y\) axis directed
  upwards.
  }
\label{fig:reactor}
\end{figure}

The geometry of the experimental configuration (see Figure \ref{fig:reactor})
consists of a bed built out of 21 alternating layers of spheres arranged in the
BCC packing. The spheres are stacked in such a manner that the odd-numbered
layers comprise nine spheres, and the even layers, referred to as weak layers,
four. The sphere diameter is $D_p = \SI{0.04}{m}$. The theoretical porosity of
the BCC packing is equal to $\phi_t = 0.32$.

The origin of the experimental coordinate system coincides with the baseplate's
top surface as indicated in Figure \ref{fig:base_plate}, which also depicts
nine bearings (spherical insets) manufactured in the base plate to hold the
lowest layer of spheres. This placement positions the centres of the spheres in
the first layer at \(y= 0.475D_p\). The plate is perforated with 312 gas flow
inlet holes, with a diameter of \(D_h = 0.1D_p\) each.

The assembly is housed in a prismatic casing as shown in
Figure~\ref{fig:reactor_geometry} with a square cross-section. As the spheres
are touching the casing walls, its size can be determined from the geometry of
the BCC packing and is given as \(D_p(4\sqrt{3}/3 + 1)=\SI{0.132}{m}\). The bed
height with all 21 layers of spheres in place is \SI{0.5}{m} as the centres of
the spheres in the \nth{21} layer are placed at \(y = 12 D_p\). The test
chamber has a height of \SI{0.92}{m} to reduce the influence of the environment
on the freeboard. 

For a direct comparison between the measurements and the numerical predictions
differences between the ideal CFD model and the geometry in the experiment have
to be considered. These differences occur due to manufacturing processing on
the side of the experiment. For the test case considered in this paper, the
spherical particles are glued together. The gluing process creates a bridge
between the individual spheres, leading to a geometry deviating from the ideal
BCC packing, increased vertical dimension of the whole particle assembly, and
slightly increased size of the inter-particle gap in the interior of the bed.
The PIV data presented in the paper has been appropriately shifted to account
for this change in size. However, these features could not be exactly
reproduced in the numerical model, which, therefore, is a high-fidelity but
nevertheless approximate representation of the experimental configuration.

The present work considers two cases corresponding to two experimental
data sets, characterised by Reynolds number based on the particle size
\(\mathrm{Re}_p = 300\) and \(500\). The Reynolds number is defined here as
\begin{equation}\label{eq:re_def}
\mathrm{Re}_p = \dfrac{\rho D_p \langle v \rangle}{\mu},
\end{equation}
where \(\mu = \SI{1.882e-5}{Pa\, s}\) is the air viscosity at the ambient
conditions (pressure of \SI{1}{bar}, the temperature of
\SI{20}{\degreeCelsius}) and \(\langle v \rangle\) is the intrinsic average of
the velocity of the gas flowing through the packing \citep{whitaker1999}.

The intrinsic average of velocity employed in Equation \eqref{eq:re_def} can be
readily obtained from the volumetric flow rate measured in the experiments as
\(\phi_t\langle v \rangle = Q / A\), where \(Q\) is the flow rate and \(A\) is
the cross-section of the packing. The flow rates for \(\mathrm{Re}_p = 300\)
and \(500\) are given as \(38.2\) and \(63.6\; \mathrm{l}\,\mathrm{min}^{-1}\)
respectively.

\section{Mathematical model}
\label{sec:math-model}

\begin{figure*}
  \subfloat[]{%
    \label{fig:com_geo}
    \includegraphics[height=7cm]{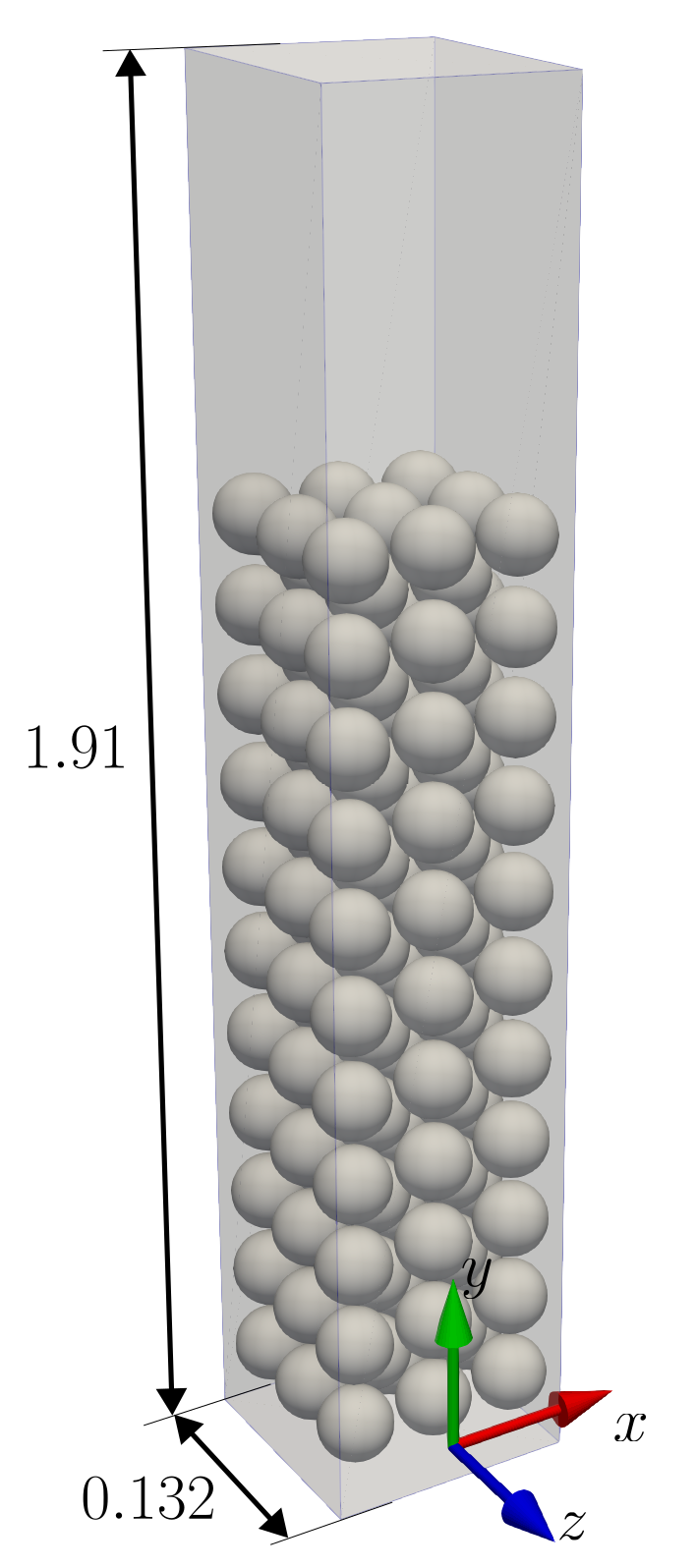}
  }\qquad
  \subfloat[]{%
    \label{fig:mesh}
    \includegraphics[height=7cm]{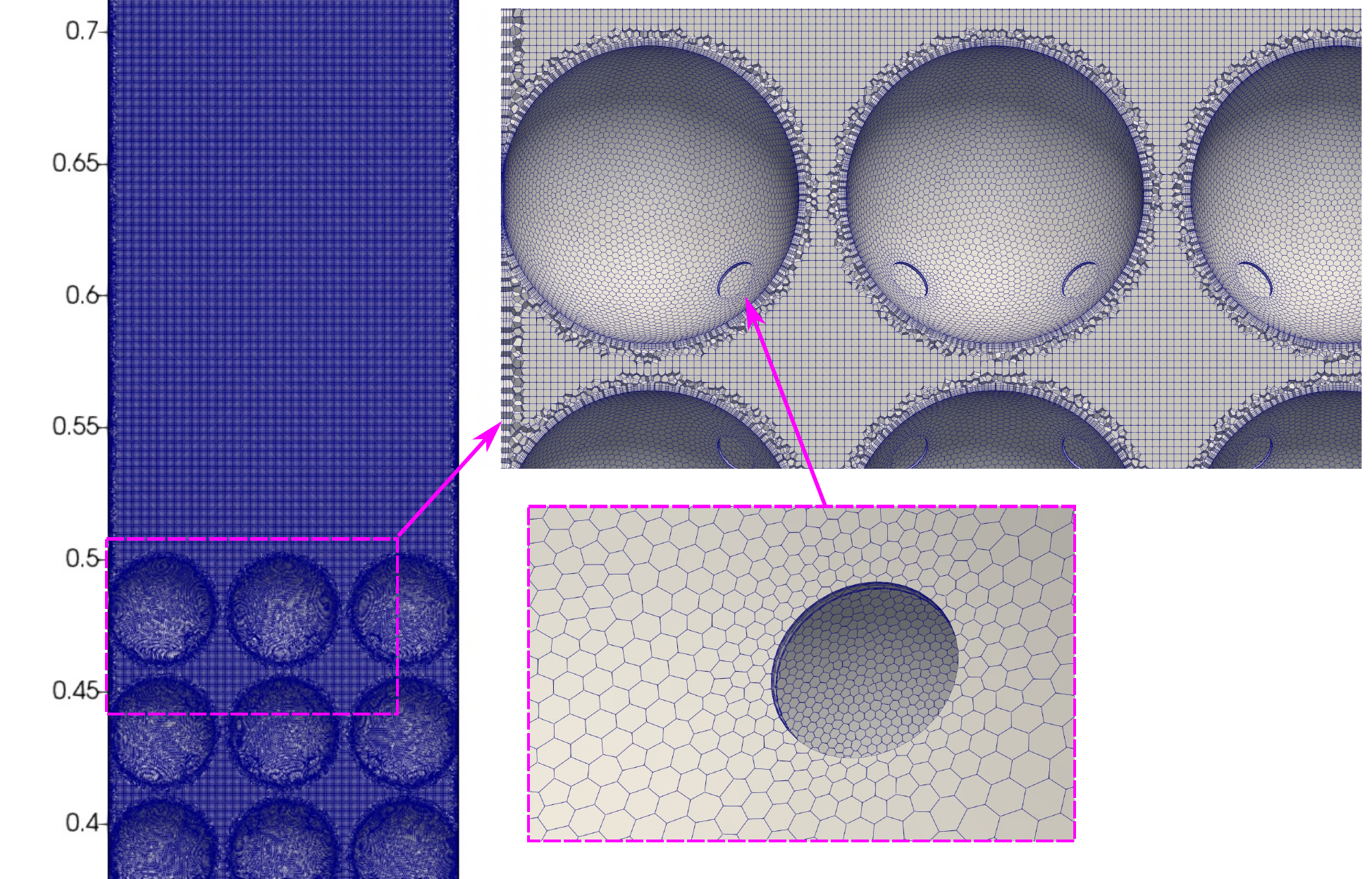}
  }
  \caption{Left figure: schematic of the computational domain (the height
  dimension is not to scale). Right picture: computational mesh clipped at the
  symmetry plane perpendicular to the \(z\) axis. The zoomed-in view shows
  the mesh around the simplified contact point between spheres in two different
  layers.}
  \label{fig:geo_and_mesh}
\end{figure*}

The simulations presented in the study are performed with the pimpleFoam solver
from OpenFOAM, a FVM-based CFD package (version v2306). For a thorough
description of numerical methods used in OpenFOAM, the reader is referred to
the works by \citet{jasak1996} and \citet{moukalled2016}. The solver is based
on the low-Mach formulation of the Navier--Stokes equations
\begin{align}\label{eq:incompressible_ns_mom}
  \pdv{u_i}{t} + u_j\pdv{u_i}{x_j} &= -\dfrac{\partial}{\partial x_i}\left(\dfrac{p}{\rho}\right)
  + \nu\dfrac{\partial^2 u_i}{\partial x_j \partial x_j},\\
  \pdv{u_i}{x_i} &= 0,
\end{align}
where \(\bm{u}\) is the velocity, \(p\) pressure and \(\nu = \mu/\rho\) is the
kinematic viscosity. In their work, \citet{gorges2024comparing}  conducted
computations of the same configuration without a sub-grid scale model and
\citet{komen2017}  successfully employed a similar setup. To assess the
sensitivity of the predictions presented in this paper to the turbulence model,
precursory calculations were performed with and without the sub-grid scale
model and the influence of the latter on the results was indeed found to be
negligible for the mesh resolution and Reynolds numbers considered. Hence, we
opted not to employ any turbulence closure for the presented computations. The
computational domain is shown schematically in Figure \ref{fig:com_geo}. The
coordinate system of the numerical domain is defined so that it coincides with
the one used to describe the experimental rig.

To simplify the meshing process, the baseplate, along with the honeycomb
structure, have been replaced with an idealised planar inlet placed at \(y =
-D_p\) (see Figure~\ref{fig:geo_and_mesh}). The lack of the baseplate in the
meshed geometry allows for high-quality cells near the bottom of the \nth{1}
layer of spheres, where highly skewed cells are generated near the
bearings, which could influence the stability of the computation.

Following the observations made by \citet{velten2023} and \citet{neeraj2023a},
in the studied configuration, the flow in the interstices and above the bed
seems to be independent of the bed height, provided the particle assembly is
made up of more than 19 layers. This effectively rules out any influence of the
flow conditions at the inlet on the unsteady flow at the freeboard, which
justifies the simplification of inlet geometry.

The section of the structure downstream of the spheres has also been extended to
minimise the influence of the outlet condition on the unsteady flow in the
freeboard. Hence, in the computations, the domain has dimensions of
\(0.132 \times 0.132 \times \SI{1.91}{m}\).

In the context of particle-resolved simulations on boundary-conforming meshes,
some geometry simplifications are often required to avoid numerical issues and
time-step limitations caused by small and often malformed cells generated by
the meshing algorithm at the contact point of two particles. \citet{dixon2020}
and references therein recommend the so-called \emph{bridges} method, which is
based on connecting the abutting geometries with a cylindrical solid volume.
The diameter of the cylinder should not be greater than \SI{20}{\percent} of
the particle diameter. This approach has been used in the current study for the
contact points between the spheres themselves and the areas between the
spheres and the casing walls. It also represents well the fact
that spheres have been glued. The cylinder diameter used for the bridges was
set to \(0.015 D_p\). Figure \ref{fig:mesh} shows the mesh around the
simplified geometry.

Typically, when conducting scale-resolving simulations, the preferred meshing
strategy consists of preparing a high-quality, smooth, preferably structured mesh
with hexahedral elements. However, such an approach is challenging to
implement in complex geometries like the one studied in the present work and, if
used, would lead to exceedingly high cell counts. Instead, a hybrid
poly-hexacore mesh was generated using the Ansys Fluent meshing module\cite{ansysfl}, due to the
ability of the meshing algorithm to generate unstructured polyhedral mesh near
the solid boundaries, while the core flow is still resolved on a uniform
hexahedral grid. The resulting mesh is characterised by low non-orthogonality and
skewness, which makes the computational method more robust.
The mesh is visible in Figure \ref{fig:mesh} and consists of \SI{19}{M} cells, of which around \SI{51}{\percent} are characterised by polyhedral shapes.
The statistics of the mesh are listed in Table \ref{tab:mesh}.

\begin{table}[t]
  \centering
  \caption{Statistics of the mesh employed in the computations, face
  non-orthogonality, cell skewness and non-dimensional wall spacing computed for
  each simulation.}\label{tab:mesh}
  \resizebox{\columnwidth}{!}{%
    \begin{tabular}{@{}llllllllllll@{}}
    \toprule
    \multicolumn{2}{c}{Non-ortho.}                & \multicolumn{2}{c}{Skewness}                 & \multicolumn{4}{c}{$y^+$}                                                 & \multicolumn{4}{c}{$x^+/z^+$}                                                 \\\cmidrule(lr){5-8}\cmidrule(lr){9-12}
    \multirow{2}{*}{Avg.} & \multirow{2}{*}{Max}  & \multirow{2}{*}{Avg.} & \multirow{2}{*}{Max} & \multicolumn{2}{c}{$\mathrm{Re}_p = 300$}        & \multicolumn{2}{c}{$500$}                 & \multicolumn{2}{c}{300}               & \multicolumn{2}{c}{500}               \\\cmidrule(lr){5-6}\cmidrule(lr){7-8}\cmidrule(lr){9-10}\cmidrule(lr){11-12}
                          &                       &                       &                      &  Avg.  &  Max  & Avg.      &  Max  & Avg.       &  Max & Avg.    &  Max \\ \midrule
      \SI{4.5}{\degree}     & \SI{58.9}{\degree}  & \num{0.1}             & \num{3.7}            &  \num{0.1}            &   \num{0.6}      &  \num{0.17}  & \num{0.75}        &  \num{1.5}  &  \num{9.5} & \num{2.15} &   \num{11.5}                \\ \bottomrule
    \end{tabular}
  }
\end{table} 

Polyhedral meshes were previously successfully used for scale-resolving
simulations of packed beds
\citep{shams2013,shams2013b,shams2013c,shams2014_stacked}. Their performance in
OpenFOAM for scale-resolving simulations, along with other mesh types, has also
been rigorously tested in a fully turbulent pipe flow \citep{komen2014},
showing a satisfacorily agreement with results generated by highly accurate
spectral codes.

\begin{figure*}[t]
  \begin{center}
    \includegraphics[width=0.8\textwidth]{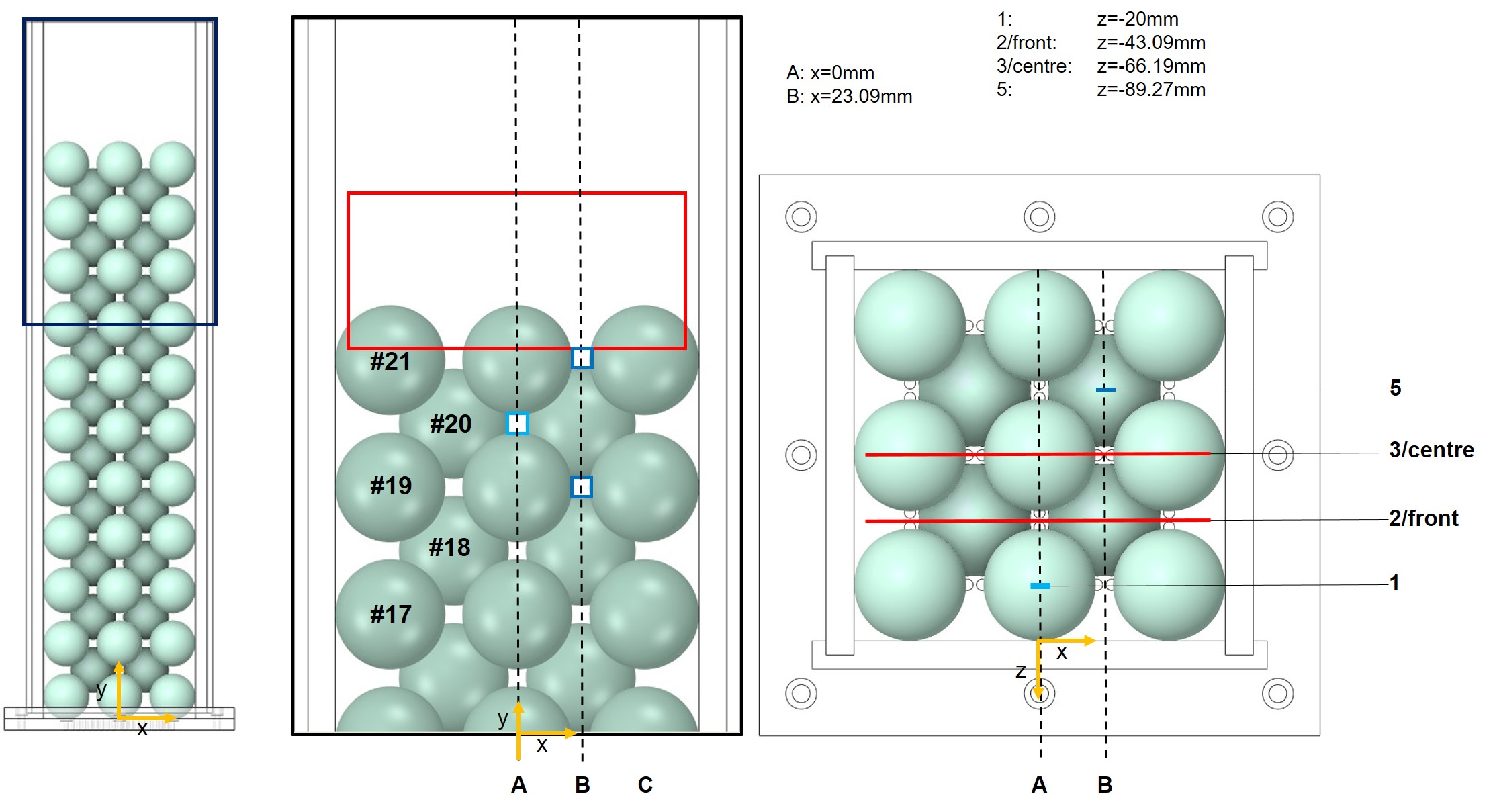}
  \end{center}
  \caption{
  Front view of the packed bed configuration with 21 layers of spheres in a
  BCC packing (left). The black rectangle marks the zoomed-in region into the layers of interest 
  (17-21), visible in the centre. In this view, the
  measurement region above the bed is marked as a red rectangle. The small blue squares
  locate the flow data
  inside the packed bed. Here, the dark blue marks the regions of interest for
  the full layers (19 and 21) and the light blue represents the interstices in
  layer 20. In the top view (right), red lines show the \(z\) coordinate of the four measurement 
  planes: 1, 2 (or front plane), 3 (or centre plane) and plane 5.
  }
  \label{fig:OverviewPosition}
\end{figure*}
  
The cubical cells in the core flow have a size of approximately \(D_p/40\),
which should provide sufficient resolution for the selected simulation,
considering that \citet{gorges2024comparing} were able to adequately resolve
most of the mean features of the flow using IBM and a grid with spacing equal
to \(D_p/32\). To resolve the contact regions appropriately, the cell size has
been set to \(D_p/200\) in these regions. The boundary layer is resolved with 4
layers of prismatic cells with the expansion rate of \num{1.2}. The region of
the mesh above the outlet of the experimental rig
\(y>\SI{23}{D_p}=\SI{0.92}{m}\) has been meshed by extruding the top boundary,
generating prismatic cells with a polyhedral cross-section. The mesh has been
coarsened in the streamwise direction \(y\) to generate a sponge layer the
outlet and thus further prevent back propagation of disturbances upstream. 

Both simulations were performed using a second-order central scheme for the
convective terms along with second-order accurate interpolation schemes. For
time integration, a second-order implicit backward scheme was used. The PISO
(Pressure-Implicit Splitting Operator) algorithm\cite{issa1986} was adopted
with five corrector loops and three sub-iterations to correct for
non-orthogonality. In the case of \(\mathrm{Re}_p=300\), the time step was
chosen to be \(\Delta t = \SI{0.0008}{s}\), whereas for \(\mathrm{Re}_p=500\)
it was equal to \(\Delta t = \SI{0.0005}{s}\), resulting in a CFL number
smaller than 0.8 in both cases.

Uniform velocity was prescribed as an inlet condition with a value computed
from the volumetric flow rates measured during the experiments. The reference
value for pressure has been prescribed at the outlet, while, for velocity, a
surface normal gradient equal to zero has been assumed. The surfaces of the
bridging cylinders are treated as walls with no-slip conditions for
velocity.

With the flow through time defined using the size of the packing
\(L_z=\SI{23}{D_p}=\SI{0.92}{m}\) and the intrinsic velocity in the packing:
\(T_\mathrm{FFT} = L_z / \langle v \rangle \), to allow for a full development
of the flow field above the bed before the collection of the statistics, both
simulations were run for \(4.5T_\mathrm{FFT}\) and \(7T_\mathrm{FFT}\), for
\(\mathrm{Re}_p=300\) and \(500\) respectively. Following that, results for
\(\mathrm{Re}_p=300\) and \(\mathrm{Re}_p=500\) were ensemble-averaged for
\(37T_\mathrm{FFT}\) and \(83.5T_\mathrm{FFT}\). These values correspond to
around \SI{300}{s} and \SI{405}{s} of simulated flow time.

\section{PIV setup}
\label{sec:piv-setup}

The air flow entering the rig is regulated by a Bronkhorst Mass-Stream
Controller (D-6371; \SI{500}{l/min} air) and seeded with Di-Ethyl-Hexyl-Sebacat
(DEHS) particles provided by a nebulizer. A Quantel Q-smart Twins 850 Nd:YAG
PIV-laser and a light sheet optic is used to create a laser light sheet and
illuminate the tracer particles in the plane of interest, which can be reached
by moving the packing on the 3D traversing system (see
Figure~\ref{fig:reactor_geometry}). For measurements above the bed a LaVision
Imager LX 8M camera equipped with a Nikon AF Nikkor \SI{35}{mm} f/2D lens was
used. The interstitial flow was captured with a LaVision Imager Intense camera
equipped with a Nikon Micro-Nikkor \SI{105}{mm} f/2.8D lens.

Image recording took place with a frequency of \SI{2.5}{Hz} and the
recording time of \SI{120}{s} and \SI{400}{s} (300 and 1000 recorded
double frame images) for \(\mathrm{Re}_p = 300\) and \(500\) respectively.
The double frame images are evaluated by a cross correlation method using Davis
8.4.

The measurement regions for the data used in this work are illustrated in
Figure~\ref{fig:OverviewPosition}. The region above the bed is
investigated in two different planes: front ($z = -\SI{43.09}{mm}$) and centre
($z = -\SI{66.19}{mm}$) marked by a red rectangle in the front view (middle
part of Figure~\ref{fig:OverviewPosition}) and red lines in the top view (on
the right). For the interstitial flow, the data is acquired
for layers 19-21 and the positions are marked in blue. For better comparability to
the original data by \citet{velten2023}, the same convention is used to name the 
regions of interest: a letter denoting the \(x\) coordinate (``A'' or 
``B'' from the centre part of Figure~\ref{fig:OverviewPosition}) followed by the plane number
a backslash and the layer number, e.g., A1/20 for the light blue 
square in the middle part of Figure~\ref{fig:OverviewPosition}.

\begin{figure*}[t]
\centering
\subfloat[]{%
    \includegraphics[height=10cm]{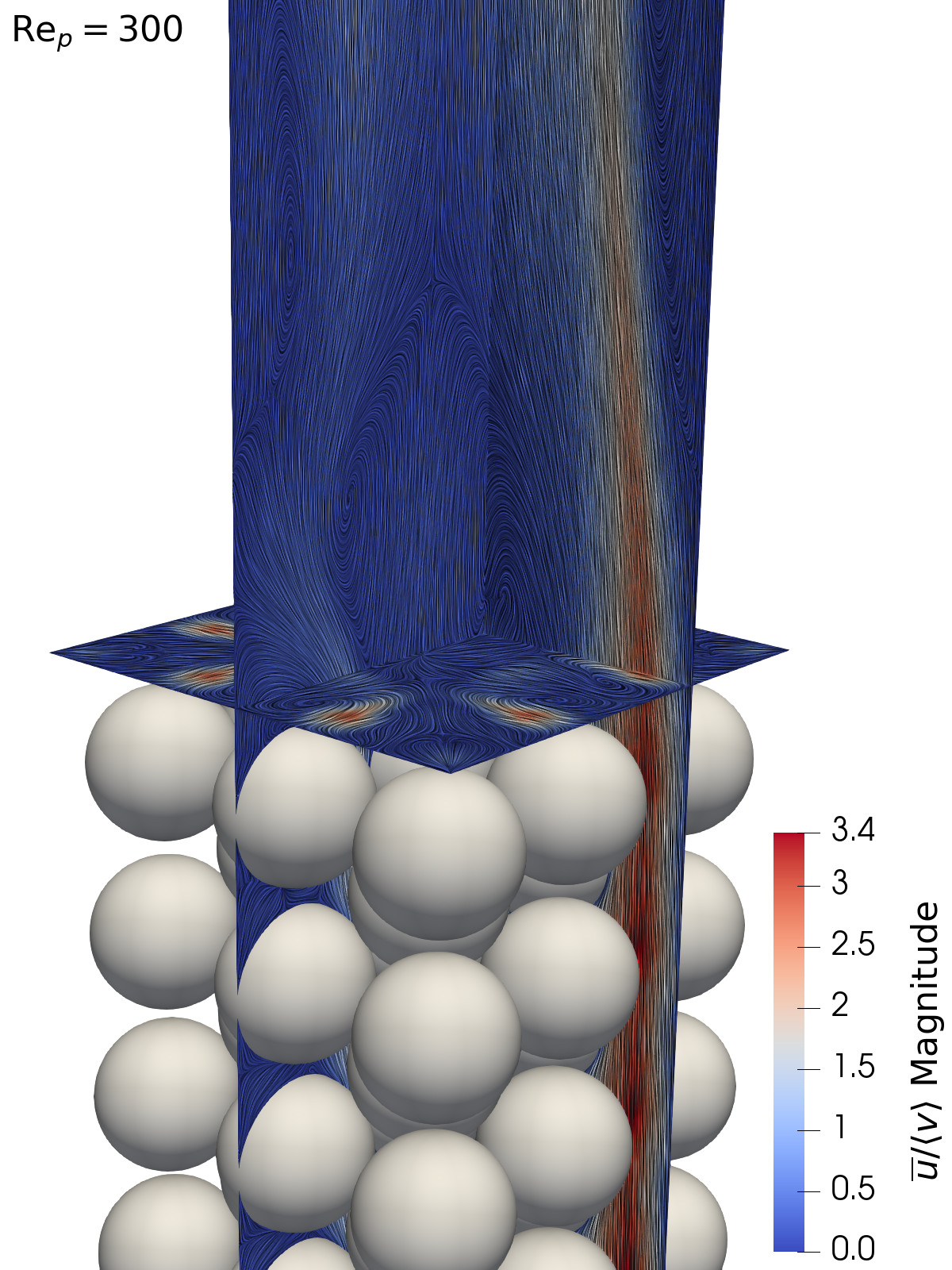}
}\quad
\subfloat[]{%
    \includegraphics[height=10cm]{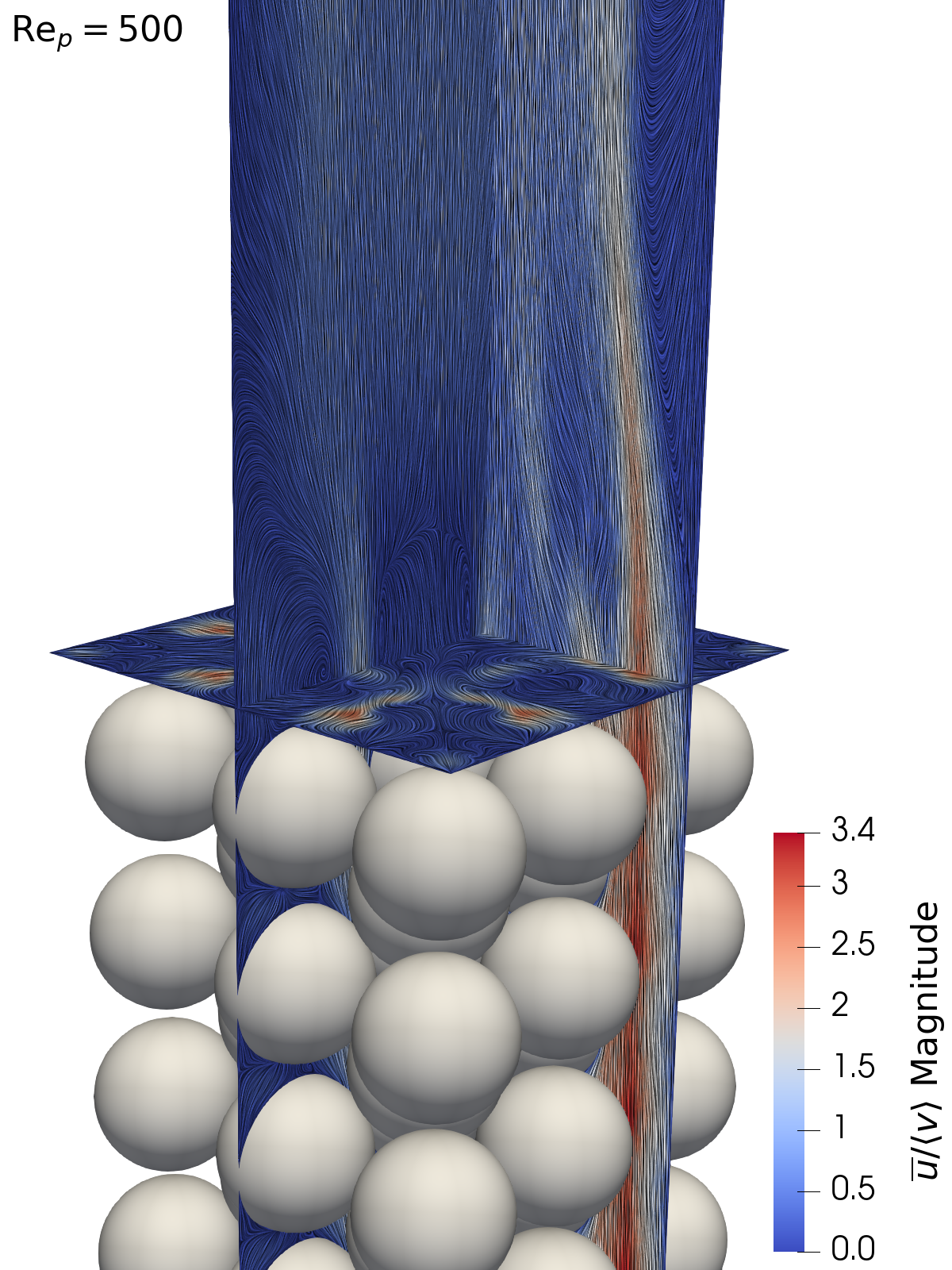}
}
\caption{Overview the flow field inside and over the studied packed-bed rector
  at \(\mathrm{Re}_p=300\) (left) and \(\mathrm{Re}_p=500\) (right). Both
  pictures illustrate the normalised ensemble-averaged velocity magnitude along
  with the streamlines visualised by Line Integral Convolution method.}
\label{fig:res_overview}
\end{figure*}

\section{Results}
\label{sec:results}
\subsection{General description of the flow field}
An impression of the computed flow field is provided in Figure~\ref{fig:res_overview},
where the mean velocity distribution is presented on the front plane, the
symmetry plane of the geometry perpendicular to the \(x\) axis and the slice at
\(y=13D_p\). The velocity distribution in the interstices, discussed at length
in Section~\ref{sec:inside}, is characterised by a strong dominance of the
wall-channeled flow. The presence of the weak layers, i.e., the lack of the
spheres near the walls in the even layers, enables an unobstructed flow, which
has a strong influence on the velocity field downstream in the freeboard.

The fluid exits the packed with non-uniform velocity; fluid in
the centre of the geometry is noticeably slower then the wall-channelled fluid,
resulting in the presence of large recirculation regions over the surface of
the packing (see Section~\ref{sec:above} for detailed description).
Additionally, the jets of fluid issuing from the near-wall interstices are directed towards the centre of the packing. This process is visibly faster for
the higher \(\mathrm{Re}_p\), due to the presence of secondary jets which contribute
to a redistribution of the momentum and thus tend to homogenise the upwards flow.
Whilst essentially stationary conditions prevail inside the bed, the freeboard flow
is dominated by low-frequency swirling and flapping motions, as also pointed
out by~\citet{neeraj2023a}. The regions of unsteady flow, discussed in 
Section~\ref{sec:unsteady} are located around the jets, however the onset of 
fluctuating motion can be identified below the crests of the topmost spheres.

\subsection{Flow inside the packed-bed}
\label{sec:inside}

\begin{figure}[b]
  \begin{center}
    \includegraphics[width=\columnwidth]{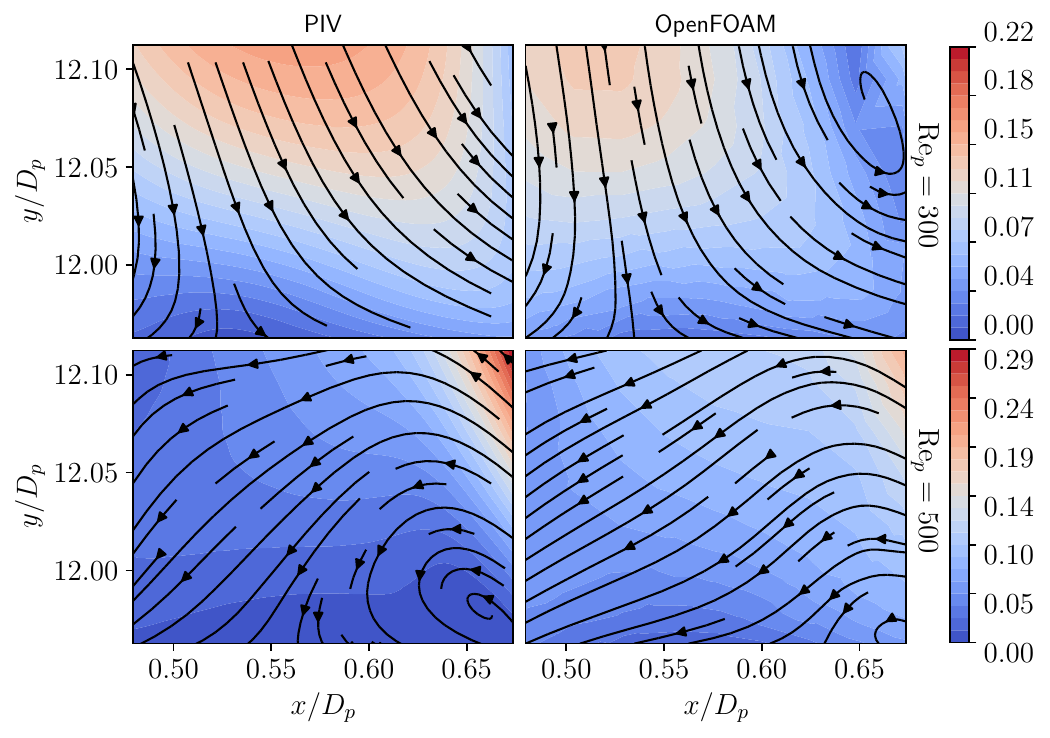}
  \end{center}
  \caption{%
  The magnitude of normalised ensemble-averaged velocity \(|\overline{\bm{u}}/\langle v \rangle|\):
  a comparison between the PIV measurements (left) and the results of
  the simulation (right) for $\mathrm{Re}_p=300$ (top row)
  and $500$ (bottom) in region B5/21 from Figure~\ref{fig:OverviewPosition} (see also annotations in 
  Figure~\ref{fig:odd_comp}). Arrows denote streamlines of the velocity field. 
  Mind the different colour bars for different \(\mathrm{Re}_p\).
  }
  \label{fig:B2_L21_vel_mag}
\end{figure}

\begin{figure}[t]
  \begin{center}
    \includegraphics[width=\columnwidth]{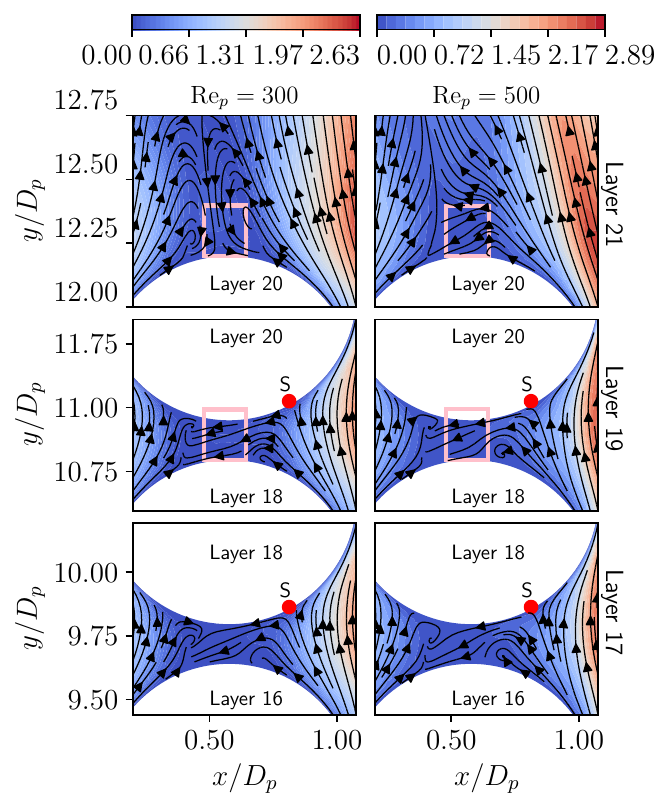}
  \end{center}
  \caption{The magnitude of normalised velocity \(|\overline{\bm{u}}/\langle
  v \rangle|\) in the interstices of the \nth{17}, \nth{19} and
  \nth{21} layer for $\mathrm{Re}_p=300$ (left) and $500$ (right), plotted at
  the plane 5 (see Figure~\ref{fig:OverviewPosition}). Mind the different colour 
  bars for different \(\mathrm{Re}_p\).
  Arrows denote streamlines of the velocity field. Label ``S'' indicates
  schematically the position of the stagnation point discussed in Section~\ref{sec:inside}. 
  The boxes show the regions measured by PIV, B5/21 and B5/19.
  }
  \label{fig:odd_comp}
\end{figure}

\begin{figure}[t]
  \begin{center}
    \includegraphics[width=\columnwidth]{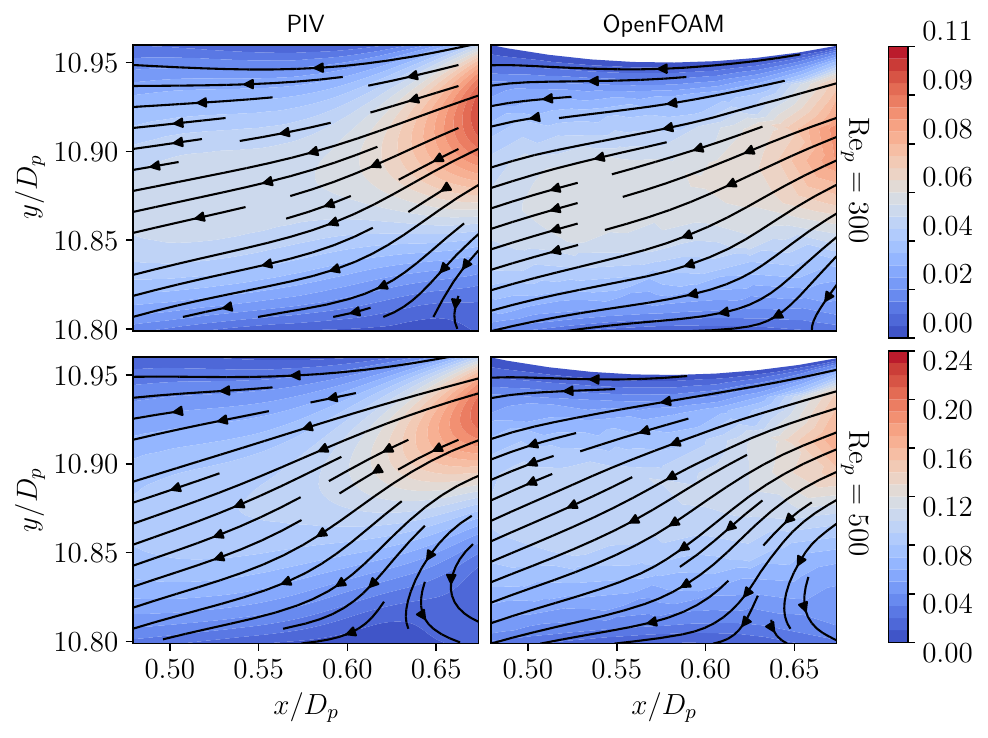}
  \end{center}
  \caption{
  The magnitude of normalised ensemble-averaged velocity \(|\overline{\bm{u}}/\langle v \rangle|\):
  a comparison between the PIV measurements (left) and the results of
  the simulation (right) for $\mathrm{Re}_p=300$ (top row)
  and $500$ (bottom) in region B5/19 from Figure~\ref{fig:OverviewPosition} (see also annotations in 
  Figure~\ref{fig:odd_comp}). Arrows denote streamlines of the velocity field. 
  Mind the different colour bars for different \(\mathrm{Re}_p\).
}
  \label{fig:B2_L19_vel_mag}
\end{figure}

Focusing first on the interstitial flow, in Figures~\ref{fig:B2_L21_vel_mag}
and~\ref{fig:B2_L19_vel_mag}, we compare the measured and simulated velocity
fields inside the bed, in the interstices near the wall.
Additionally, to further illustrate the
features of the flow, in Figure~\ref{fig:odd_comp}, we discuss the properties
of the velocity field around the measurement sections (the regions captured by
PIV are also indicated in the same figure).

Figure~\ref{fig:B2_L21_vel_mag} shows the recirculation region near the surface
of the bed above the spheres in the \nth{20} layer for both
\(\mathrm{Re}_p\). The most pronounced feature of the flow in the neighbouring
area is the fast-moving jet on the right hand side, accelerated in the openings
between the packing and the wall, and released into the
freeboard via the interstices of the \nth{21} layer. These fast-moving jets,
for both \(\mathrm{Re}_p\) are visible in Figure~\ref{fig:odd_comp} on the
right of each plot. As indicated in the figure, the PIV measurement regions
were positioned to the left of one of the wall-adjacent interstices, i.e.\ in the
areas exhibiting high flow velocity due to the wall-channelling
effects. This in turn leads to the presence of a large recirculation region
above the spheres in \nth{20} layer visible in the PIV and simulation data in
Figure~\ref{fig:B2_L21_vel_mag}.

For \(\mathrm{Re}_p=300\), the measurement section is dominated by a downwards
directed flow, which slows down while approaching the surface of the sphere (located
just below the measurement window). The bulk of the flow is redirected to the
right, where it joins the fluid accelerated next to the wall. Due to the
presence of the faster flow near the wall, the velocity field is visibly
asymmetric with respect to the sphere positioned below the measurement section,
centered at \(x=0.57725 D_p\). Thus, the reattachment point at the
surface of the sphere is visibly skewed to the left side of the observed region.

The main difference between the simulated and measured flow is the presence of a
vortex in the simulation data for \(\mathrm{Re}_p=300\) on the top-right part
of Figure~\ref{fig:B2_L21_vel_mag}. This feature is not visible in the PIV
measurments. It can be argued that the recirculation region
as a whole is narrower than in the experimental measurement of the flow field.
However, the overall character of the flow is reproduced well in the simulated
results.

On the other hand, in the case of \(\mathrm{Re}_p = 500\), the comparison
between the predicted velocity field and the PIV measurement is even more satisfactory
(Figure~\ref{fig:B2_L21_vel_mag}). 
As for lower \(\mathrm{Re}_p\), the flow separates from the sphere in the
\nth{20} layer; however, the recirculation region is much smaller, with the
reattachment occurring right of the centre of the measurement section. Further,
the flow in this case is mainly directed downwards and towards the centre of
the structure, which seems to drive a faster reattachment of the separation
bubble. Similarly to the \(\mathrm{Re}_p=300\) results, the predicted separated
region is noticeably smaller than in the PIV measurement.

Overall, looking at Figure~\ref{fig:odd_comp}, for \(\mathrm{Re}_p=300\), the
recirculation region above the \nth{20} layer appears to be stretched in the vertical
direction due to the limited space between the spheres in \nth{21} layer. Yet,
the velocity field of the \(\mathrm{Re}_p=500\) case is visibly more complex.
Instead of forming a pair of vortices, the flow from the near-wall jets
impinges on the fluid coming from the left, forcing a closer reattachment of
both recirculation regions. Nevertheless, both flows share many similarities,
with the recirculation regions skewed towards the centre of the studied geometry and the overall character of the flow field being clearly dominated by the
behaviour near the walls.

Figure~\ref{fig:B2_L19_vel_mag} shows the flow conditions directly below
the previous measurement window, at the height of the \nth{19} layer of
spheres. Here, simulation results are also in very close agreement with the
experimentally measured velocity field. For both Reynolds numbers, the overall
character of the flow is adequately reproduced in the simulated data.

For each even layer, the flow next to the walls impinges onto the
surface of the higher spheres due to the expansion of the interstitial space
next to the wall (layers 17 and 19 at the right sides of the colour maps in
Figure~\ref{fig:odd_comp}, stagnation point schematically indicated by the letter S). Part of the fluid is
then diverted into the narrow gap between the spheres in the odd layers. This
faster moving swath of fluid enters the measurement area in the top-right
corner for both \(\mathrm{Re}_p\) numbers. This results in the flow being
directed towards the centre of the geometry and the presence of a recirculation
and reattachment in the bottom part of the region of interest. The flow
reattaches on the sphere in the next layer and part of the fluid returns to the
``channel'' near the wall. For \(\mathrm{Re}_p= 500\), faster fluid is
transported deeper towards the centre of the structure, increasing the size of
the separation bubble w.r.t.\ lower \(\mathrm{Re}_p\).
Inspection of the velocity field on the other side of the area captured
by the PIV data and illustrated in Figure~\ref{fig:odd_comp} also reveals a
complex behaviour. The fluid channelled between the spheres in even layers mixes
with the vertical flow convected through the middle of the packing, resulting in
a stronger recirculating flow. 

Figure~\ref{fig:odd_comp} shows clearly how the proximity of the bed surface
influences the interstitial flow field in the top three odd layers.
In between the layers \nth{16} and \nth{18}, as well as \nth{18} and \nth{20},
the velocity fields remain remarkably similar. On the other hand, the
differences in the flow field in the interstices of the \nth{19} and \nth{21}
layers are much more pronounced. This can be explained by the fact that the
flow is able to separate effectively with no constraints from the \nth{20}
layer of spheres.

This seems to suggest that the conditions near the surface of the packing have a
strong influence on the flow inside the last layer of spheres, but not further
downwards, which appears to be mostly driven by the geometry and arrangement of
the particles. For the BCC packing under the investigated Reynolds numbers, the
main features of the flow in the freeboard are therefore mostly determined by
the arrangement of the particles in the top two layers. Additionally, when no
backflow from the freeboard into the bed is present, the flow conditions deviate from the ones inside the packed bed only
near the two topmost layers.

\begin{figure}
  \begin{center}
    \includegraphics[width=\columnwidth]{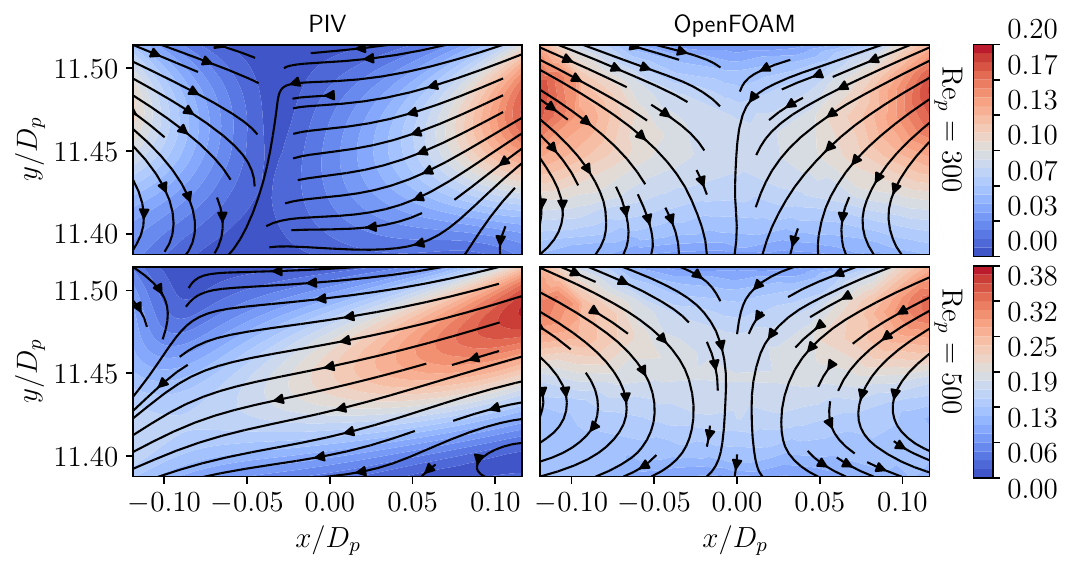}
  \end{center}
  \caption{
  The magnitude of normalised ensemble-averaged velocity \(|\overline{\bm{u}}/\langle v \rangle|\):
  a comparison between the PIV measurements (left) and the results of
  the simulation (right) for $\mathrm{Re}_p=300$ (top row)
  and $500$ (bottom) in region A1/20 from Figure~\ref{fig:OverviewPosition}. 
  Arrows denote streamlines of the velocity field. 
  Mind the different colour bars for different \(\mathrm{Re}_p\).
  }
  \label{fig:A8_L19_vel_mag}
\end{figure}

Figure \ref{fig:A8_L19_vel_mag} illustrates the flow around the \(x=0\)
symmetry plane of the structure at the height of the \nth{20} layer. In this
region, the flow field should be symmetric; this property is, however, not
fully captured in the measured data. Based on the simulation results, the
velocity should be characterised by two recirculation zones, driven by the mean
streamwise flow through the packing, meeting in the middle of the region, i.e.\
around \(x=0\). In the experimental results, for both Reynolds numbers, both
recirculation zones seem to be shifted towards the negative values of \(x\).
This effect is more pronounced for the higher \(\mathrm{Re}_p\), as the whole
measurement region seems dominated by the flow in the left direction. This
behaviour is potentially caused by the unavoidable inaccuracies resulting from
assembling and gluing the spheres in the packed bed.

\subsection{Flow above the bed}
\label{sec:above}

\begin{figure}
  \begin{center}
    \includegraphics[width=\columnwidth]{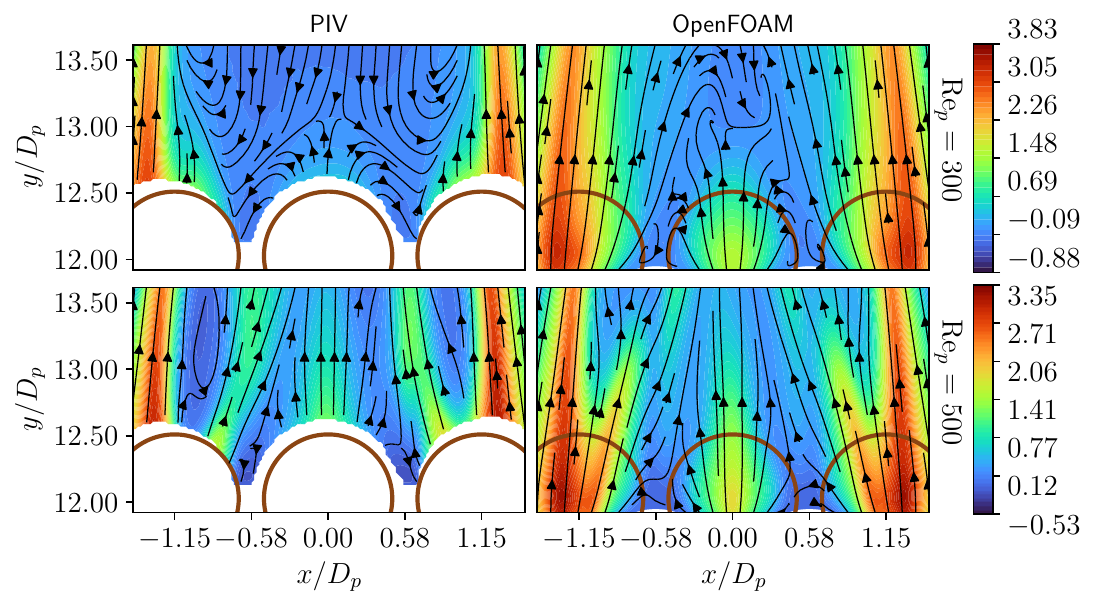}
  \end{center}
  \caption{The magnitude of vertical component of ensemble-averaged velocity 
  \(\overline{v}/\langle v \rangle\):
  a comparison between the PIV measurements (left) and the results of
  the simulation (right) for $\mathrm{Re}_p=300$ (top row)
  and $500$ (bottom) above the bed surface at the front plane (see Figure~\ref{fig:OverviewPosition}). 
  Arrows denote streamlines of the velocity field. 
  Mind the different colour bars for different \(\mathrm{Re}_p\). 
  The positions of spheres in \nth{21} layer is annotated on each graph.
  In case of PIV measurements, white
  colour denotes area where the velocity was not captured by the camera.
  } 
  \label{fig:front_vel_mag}
\end{figure}

\begin{figure}
  \begin{center}
    \includegraphics[width=\columnwidth]{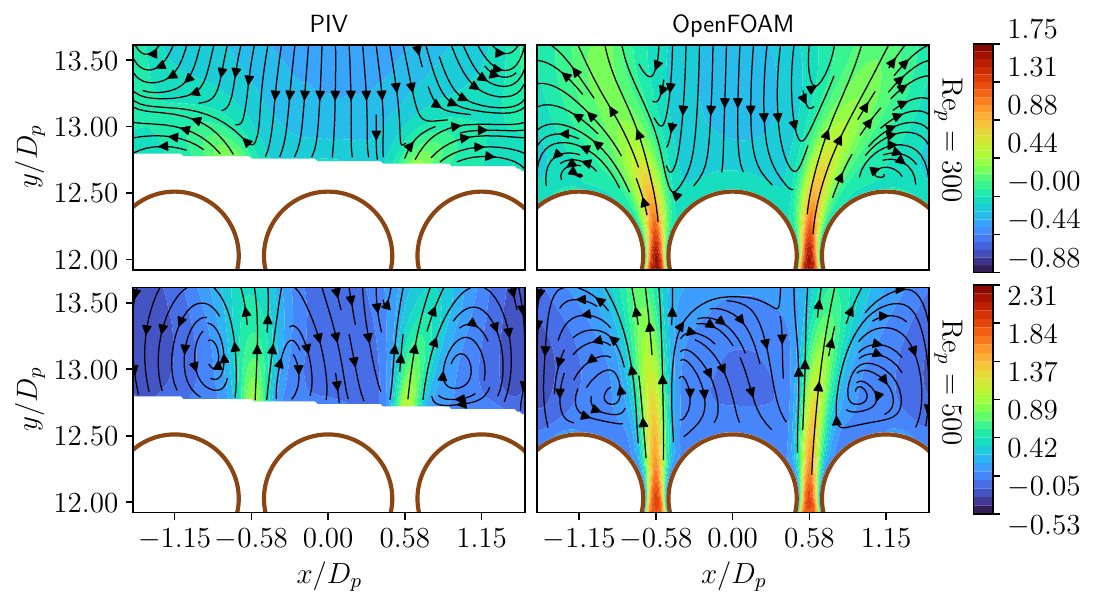}
  \end{center}
  \caption{The magnitude of vertical component of ensemble-averaged velocity 
  \(\overline{v}/\langle v \rangle\):
  a comparison between the PIV measurements (left) and the results of
  the simulation (right) for $\mathrm{Re}_p=300$ (top row)
  and $500$ (bottom) above the bed surface at the centre plane (see Figure~\ref{fig:OverviewPosition}). 
  Arrows denote streamlines of the velocity field. 
  Mind the different colour bars for different \(\mathrm{Re}_p\). 
  The positions of spheres in \nth{21} layer is annotated on each graph.
  In case of PIV measurements, white
  colour denotes area where the velocity was not captured by the camera.
  }
  \label{fig:mid_vel_mag}
\end{figure}

The ensemble-averaged, streamwise component of the velocity over the bed surface, on the
\textbf{front} and \textbf{centre} measurement planes, is presented in
Figures~\ref{fig:front_vel_mag} and~\ref{fig:mid_vel_mag} respectively. Due to
the strong wall-channelling effects, for both \(\mathrm{Re}_p\) the flow at the
front plane is dominated by two jets issuing from the bed with velocities which result to be \(3-4\) times
higher then the theoretical intrinsic average velocity \(\langle v
\rangle\). The flow originating from the bed through the gap in the centre of the
freeboard (around \(x=0\)) is much slower, with simulation results indicating
maximum velocity of around \(1.5\langle v \rangle\).

For the lower \(\mathrm{Re}_p\), the middle part of the PIV measurement section
is characterised by a backflow. Two recirculation regions are visible forcing
the fluid flowing out of the bed in the centre of the freeboard to join the flow
in the jets. This behaviour cannot be observed in the simulated results.
Qualitatively, the character of the flow field is predicted accurately by the
simulations, although the computed jets are noticeably wider and more diffused.
This in turn, seems to reduce the recirculation regions to a much smaller
vortex located over the centre of the packing, where the fluid ejected from
both wall-adjacent interstice and the one at the centre get mixed.

A similar flow structure can be observed in PIV data in the case of
\(\mathrm{Re}_p=500\). Two regions of fluid accelerated near the wall
are again the most important feature of the velocity field. Here, the simulation
results are in a very good agreement with the measured flow field. Additionally,
both measurement and simulation data indicate the presence of two additional
diagonal jets. The jets have a higher velocity near the bed in the simulations,
however, they dissipate more quickly. Additionally, no recirculation is visible
between each pair of the jets. Based on the computed velocity field, the
additional jets also originate from the interstices adjacent to the wall. The
presence of these diagonal jets seems to redistribute the velocity more evenly,
facilitating the flow leaving the bed in the middle of the packing and reducing the size of
recirculating regions of the fluid.

For both \(\mathrm{Re}_p\), the flow at the centre plane, visible in
Figure~\ref{fig:mid_vel_mag}, is characterised by the presence of two streams
ejected from the gap between the topmost layer of spheres. For
\(\mathrm{Re}_p=300\), according to the PIV results, the ejected fluid is quickly
redistributed towards the sides of the freeboard, and the jets are mostly diffused
below \(y=\SI{0.52}{m}=13D_p\). Above, in the centre, a region of
downwards directed velocity field is measured, creating two recirculation
regions above the jets.
Although the computed velocity field is in close agreement with the
measurements, higher velocities characterise the jets in the numerical predictions, which, in
turn, allows them to exist beyond \(y=13D_p\). Nevertheless,
they behave qualitatively in a similar fashion as observed in the experiments, as they are also directed towards the walls.
As the jets remain closer to the centre longer, the backflow
region above the centre of the packing is smaller and characterised by a
lower velocity. Simulation results also reveal the presence of two
recirculation regions above the surfaces of the spheres near the walls. 

As the inertial effects get stronger with higher Reynolds number, at
\(\mathrm{Re}_p=500\) the ejected fluid gets convected upwards instead of being
quickly directed towards the wall, for both experiment and simulation.  This
leads to the presence of two big recirculation regions above the two off-center
spheres and relatively strong downwards flow in the middle of the captured
region.

A small break in the symmetry of the measured velocity fields can be observed,
albeit not to a large extent: both jets are slightly angled towards the right
side of the freeboard. Similarly to the conditions inside the bed
(Figure~\ref{fig:A8_L19_vel_mag}), this was probably caused by  the already
discussed slight asymmetries of the experimental rig and packing itself. 
  
The shape of the jets are predicted much better for the higher \(\mathrm{Re}_p\),
although the computation still overpredicts the velocity in the core of each
jet. Simulation results are also not entirely symmetrical, with the middle section
in between the jets dominated by a diagonal downwards flow. The recirculation
regions resulting from the presence of the jets reattach at the surface of the
spheres.

\begin{figure}
  \begin{center}
    \includegraphics[width=\columnwidth]{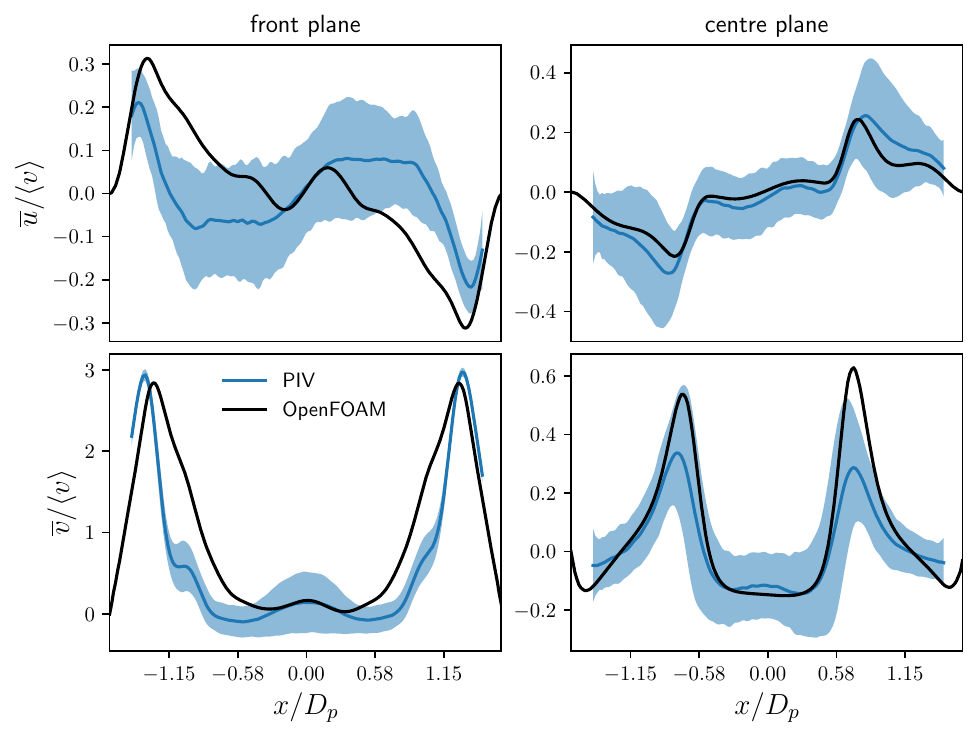}
  \end{center}
  \caption{Comparison of the ensemble-averaged horizontal (top) and vertical
  (bottom) velocity components for \(\mathrm{Re}_p=300\), at the front (left)
  and centre planes (right), plotted at \(y=12.8D_p\). The blue band denotes
  the range of values separated from the measured data by two standard
  deviations denoting the \SI{95}{\percent} confidence interval.}
  \label{fig:Re300_line_plot_comp}
\end{figure}

Figure~\ref{fig:Re300_line_plot_comp} presents a quantitative comparison of the
mean velocity components at the front and centre plane at \(\mathrm{Re}_p=300\), sampled
at \(y=\SI{0.512}{m}=12.8D_p\) from both the PIV data and simulation results. The
PIV results include a confidence interval of two standard deviations,
representing the \SI{95}{\percent} confidence level visible as the
blue band on the graphs. Focusing first on the ensemble-averaged
streamwise velocity, \(\overline{v}\), at the front plane, the jets predicted in the
simulation are noticeably wider than their measured counterparts. The peaks of
maximum velocity are also slightly shifted towards the middle of the freeboard. The PIV
results also indicate the presence of two smaller peaks in the streamwise
velocity around \(x/D_p = \pm1\), which are potentially the weaker counterparts of
the diagonal jets present in the \(\mathrm{Re}_p=500\) simulation.

The magnitude of the horizontal velocity component \(\overline{u}\)  is also
overpredicted in the same area, so that the jets predicted by the simulations are
wider due to an increased convection towards the centre of the freeboard
w.r.t.\ the PIV data set. Nevertheless, \(\overline{u}\) shows a trend similar to the measured data, with the maxima of horizontal velocity located near
the wall. It has also to be kept in mind, that the fluctuations of the
horizontal velocity component are much stronger than for the vertical one,
which can be seen from the standard deviation of the experiments. This also
influences the  mean values obtained in both experiment and simulation. 

The horizontal velocity component is in an excellent agreement with the experimental
results at the centre plane (right column of
Figure~\ref{fig:Re300_line_plot_comp}). The line plots of PIV also show a
stronger flow towards the walls then computed, identified by the higher velocity magnitude in the near-wall
regions. The streamwise velocity at the centre plane measured by PIV is also
slightly asymmetric for this lower Reynolds number. The maxima of vertical
velocity are significantly overpredicted however the overall shape and position
of the jets is rendered in a similar way to the PIV results. The recirculation
regions near the walls are also more pronounced indicating bigger recirculation
regions in the simulations.

\begin{figure}[t]
  \begin{center}
    \includegraphics[width=\columnwidth]{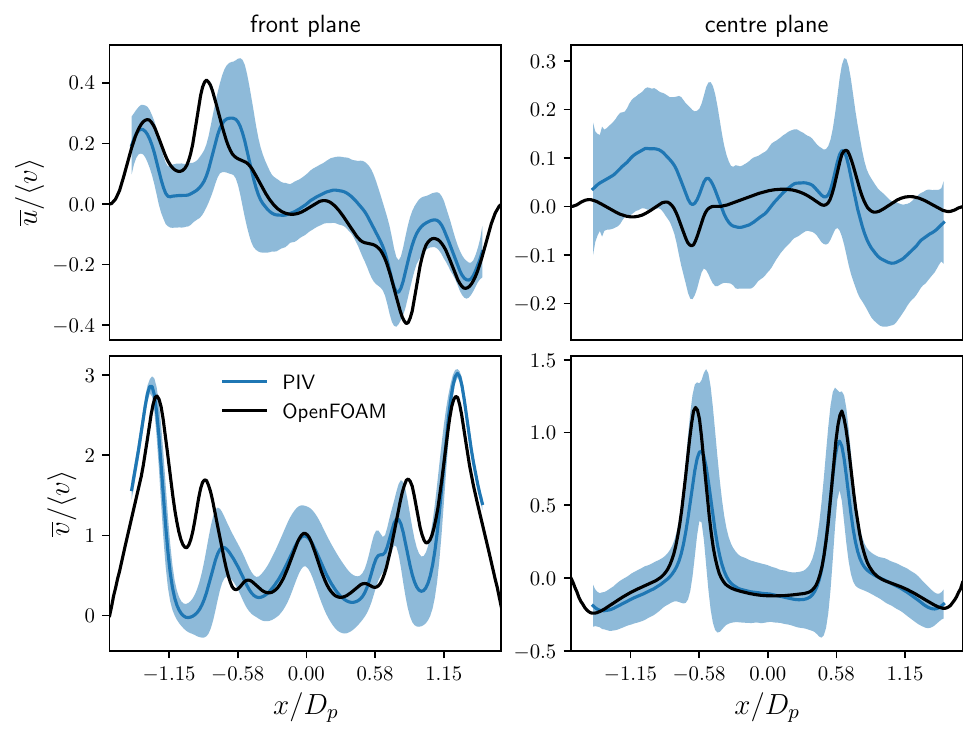}
  \end{center}
  \caption{Comparison of the ensemble-averaged horizontal (top) and vertical
  (bottom) velocity components for \(\mathrm{Re}_p=500\), at the front (left)
  and centre planes (right), plotted at \(y=12.8D_p\). The blue band denotes
  the range of values separated from the measured data by two standard
  deviations denoting the \SI{95}{\percent} confidence interval.}
  \label{fig:Re500_line_plot_comp}
\end{figure}

The same comparison was performed for the higher Reynolds number and is
presented in Figure~\ref{fig:Re500_line_plot_comp}. Focusing first on the
streamwise velocity at the front plane, the flow structure is visibly more
complex than in the \(\mathrm{Re}_p=300\) case. The velocity profile is
characterised by the presence of five peaks, corresponding to the five jets:
two ejected from the interstices near the walls as at \(\mathrm{Re}_p=300\),
two smaller ones directed towards the middle of the freeboard and the fluid
exiting the packed-bed in the centre (see the discussion related to Figure
\ref{fig:front_vel_mag}). In this case, the simulation results predict the
position and shape of the two bigger jets much better than in the case of
\(\mathrm{Re}_p=300\). Similarly to the low Reynolds number results, the
velocity in the middle is predicted to a very high degree of accruacy. The position of the
secondary jets mentioned above is visibly shifted towards the sides of the freeboard, with
much higher velocity then measured experimentally, which aligns with the
observations based on Figure~\ref{fig:front_vel_mag}. An analogous behaviour can
be observed by examining the horizontal velocity component: simulation
results match the experimental data very well, although the maxima of simulated
velocity are also shifted towards the walls.

The predicted streamwise velocity at the centre plane shows a very good agreement with
the experimental data with a slight overestimation of the maximum velocity
inside the jets and the recirculation regions near the walls are
reproduced with an excellent accuracy. The velocity prediction does not match
well  the experimental data for the horizontal component, although most of
the profiles still falls within the \SI{95}{\percent} confidence interval.
Computation predicts virtually no horizontal flow near the walls
whereas the PIV data indicates that the flow is directed towards the center of
the bed.

\subsection{Analysis of unsteady flow over the bed}
\label{sec:unsteady}

\begin{figure}
  \begin{center}
    \includegraphics[width=\columnwidth]{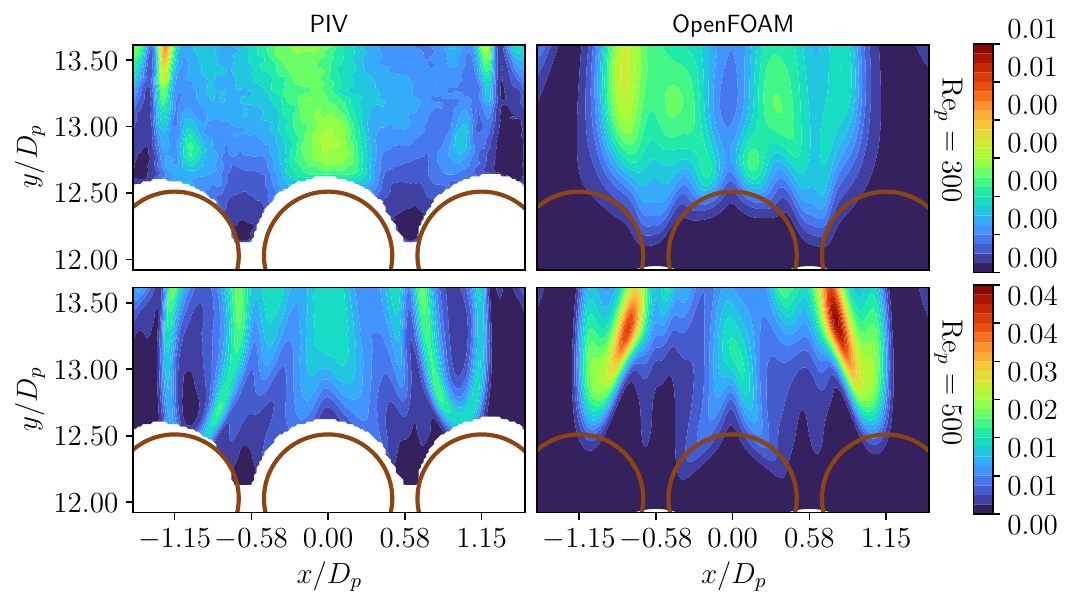}
  \end{center}
  \caption{The normalised TKE \(k/\langle v \rangle^2\):
  a comparison between the PIV measurements (left) and the results of
  the simulation (right) for $\mathrm{Re}_p=300$ (top row)
  and $500$ (bottom) above the bed surface at the front plane (see Figure~\ref{fig:OverviewPosition}). 
  Mind the different colour bars for different \(\mathrm{Re}_p\). 
  The positions of spheres in \nth{21} layer is annotated on each graph.
  In case of PIV measurements, white
  colour denotes area where the velocity was not captured by the camera.
 }
  \label{fig:front_TKE}
\end{figure}

As the flow detaches from the spheres in the topmost layer, it display a higher degree of 
unsteadiness. 
In the PIV measurement data only two components of the velocity in the
plane illuminated by the laser sheet are available, therefore the turbulent kinetic energy (TKE) is computed
in an approximated way as
\begin{equation}
  k_\mathrm{PIV} = \dfrac{3}{4} \left( \overline{u'u'} + \overline{v'v'} \right).
  \label{eq:TKE_PIV}
\end{equation}
This approximation is based on the assumption of isotropy, i.e., the third
component of normal fluctuations contains similar turbulent contribution as the
other two \cite{lavision2019}. Since in the simulations
all three components of velocity are resolved, the TKE can be computed according 
to the theoretical definition \cite{wilcox2006}:
\begin{equation}
  k = \dfrac{1}{2}\overline{u_i'u_i'}.
  \label{eq:TKE}
\end{equation}

\begin{figure}
  \begin{center}
    \includegraphics[width=\columnwidth]{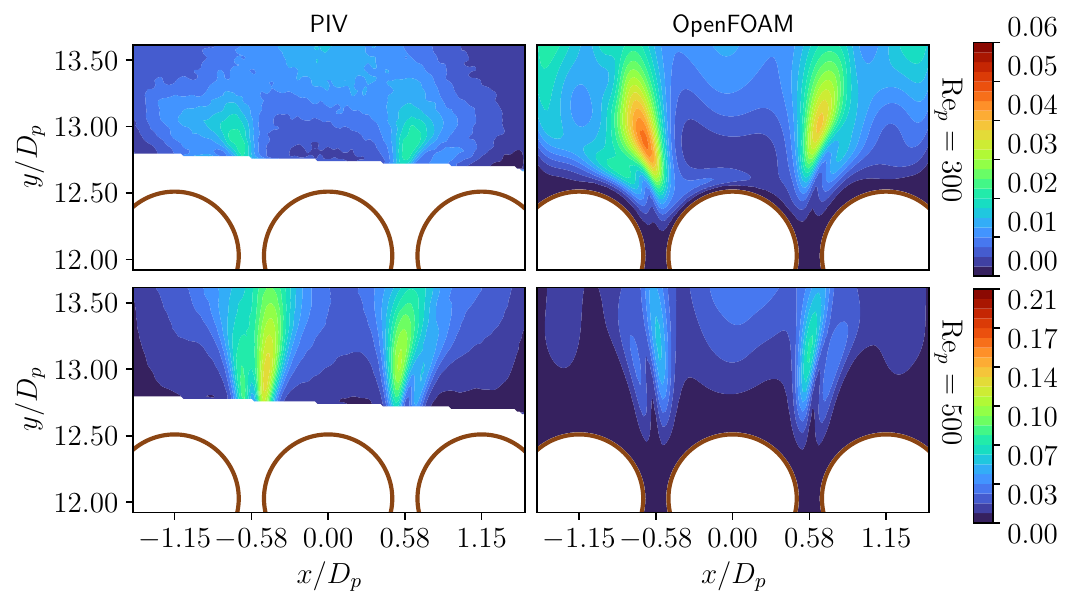}
  \end{center}
  \caption{The normalised TKE \(k/\langle v \rangle^2\):
  a comparison between the PIV measurements (left) and the results of
  the simulation (right) for $\mathrm{Re}_p=300$ (top row)
  and $500$ (bottom) above the bed surface at the centre plane (see Figure~\ref{fig:OverviewPosition}). 
  Mind the different colour bars for different \(\mathrm{Re}_p\). 
  The positions of spheres in \nth{21} layer is annotated on each graph.
  In case of PIV measurements, white
  colour denotes area where the velocity was not captured by the camera.}
  \label{fig:mid_TKE}
\end{figure}

Figure~\ref{fig:front_TKE} illustrates the distribution of the turbulent
kinetic energy (TKE) at the front plane above the bed. Data from both the PIV
measurements and simulations is presented, at both \(\mathrm{Re}_p\) numbers.
For centre plane the same comparison is visible in Figure~\ref{fig:mid_TKE}.

At the front plane, the simulation results for \(\mathrm{Re}_p=300\) are in
qualitative agreement with the data measured using PIV. There are, however,
significant differences. PIV indicates the presence of two elongated highly
unsteady regions near the walls, at the height of \(y=13.5D_p\).
Furthermore, over the surface of the bed, around \(x = \pm 1.15D_p\), the fluid
exiting the packing at the middle of the front plane and joining the flow near
the walls (see Figure~\ref{fig:front_vel_mag}) seem to create an additional
pair of small areas of unsteady flow. The middle of the freeboard is also
dominated by a region of increased TKE which gets progressively shifted to the
left side. These rather separate structures are not predicted by
the PRS, potentially due to the overly diffused jets. Instead, the TKE seems to
be located in two symmetric spots above the bed, generated both by the
influence of the shear layer due to the presence of the jets and the
oscillating flow in the middle of the front plane. The latter feature of the
PIV data is satisfyingly reproduced by the simulation results, with two regions
of increased TKE near he surface of the bed in the centre of the front plane. A conclusion on the isotropy and nature of the turbulence cannot be drawn based on the evidence of the comparison presented in the paper and further investigations are required.

For the higher Reynolds number, PIV data illustrates five distinct areas of
high TKE, corresponding to the five regions occupied by faster flow and resulting
recirculation regions, visible in Figures~\ref{fig:front_vel_mag}
and~\ref{fig:Re500_line_plot_comp}. The TKE computed from the numerical flow fiels is in a much closer agreement
to the experimental measurement then for the lower \(\mathrm{Re}_p\). The
unsteady region in the middle of the freeboard is clearly visible in the computed
field. Instead of the narrow regions of high TKE visible in the measured data
at the sides of the freeboard, the simulation predicts more diffused and bigger
swaths of unsteady fluid, with much higher values of TKE. 

At the centre plane (Figure~\ref{fig:mid_TKE}) both simulations and PIV
measurements predict remarkably similar distribution of TKE. Data captured from
the \(\mathrm{Re}_p=500\) experiment indicates the presence of four shear
layers on both sides of the jets exiting the bed via the interstices around the
sphere in the centre. The areas located on the outside of the jets have
significantly weaker TKE content, presumably due to the stabilising influence
of the near wall flow. The numerical results show a very similar pattern, however,
the value of TKE is underestimated. Although for the lower \(\mathrm{Re}_p\)
in the PIV data only two regions of high TKE are clearly visible, it can be
inferred by comparing to the computed distribution of TKE that the behaviour of
the flow field is analogous to the higher Reynolds number case. The simulation
results overpredict the value of \(k\) inside the inner shear layers, however,
it also predicts two smaller regions of increased TKE, located around the outer
sides of the jets, which get transported towards the walls by the
velocity field (see Figure~\ref{fig:mid_vel_mag}). This behaviour is in line
with what can be observed in the data available in the PIV measurements.

\begin{figure}
  \begin{center}
    \includegraphics[width=\columnwidth]{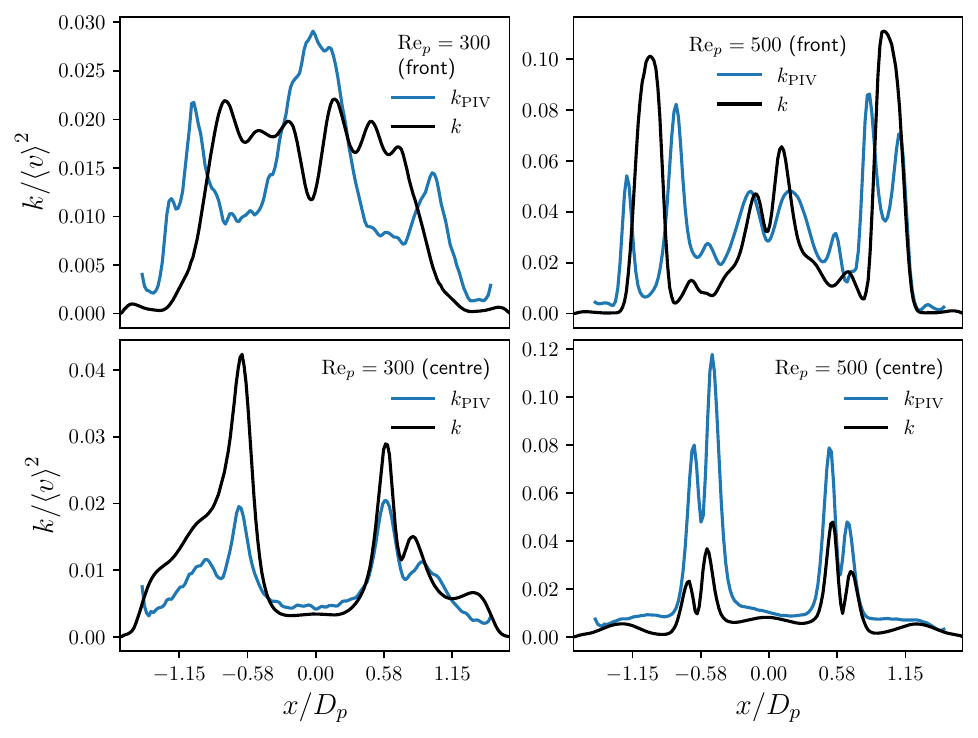}
  \end{center}
  \caption{Comparison of the normalised TKE \(k/\langle v\rangle^2\) for
  \(\mathrm{Re}_p=300\) and 500, at the front and centre planes, plotted at \(y=12.8D_p\).
 }
  \label{fig:line_TKE}
\end{figure}

The same comparison, in a quantitative way, can be made using the data
displayed in Figure~\ref{fig:line_TKE}, where the TKE is plotted along the \(y=12.8D_p\)
line in both front and centre planes. As discussed before, for
\(\mathrm{Re}_p=300\) the simulated result predict much more diffused
distribution of the TKE: the three peaks observed in the PIV data are not
present. At the centre plane, the PIV data clearly indicates the presence of
the two pairs of the shear layers as pointed out in the previous paragraph. The
simulated distribution of TKE is asymmetric with much more pronounced unsteady
behaviour near the left jet. However, the \(k\) computed for the right side of
the freeboard aligns very well with the measured data, with the peaks of TKE
overpredicted by around \SI{50}{\percent}.

For the higher \(\mathrm{Re}_p\), as mentioned previously, the distribution of
TKE is predicted very accurately in the middle of the freeboard on both
measurement planes. The high TKE regions visible in the PIV data near the
walls, at the front plane, are merged together resulting in a strongly
overestimated \(k\). As a whole, the computed TKE field clearly shares the same
features as the measured distribution. Contrary to the results obtained for the lower
\(\mathrm{Re}_p\) the value of the TKE predicted in the centre plane is mostly
underestimated, however, the peaks related to the shear layers are reproduced
well in the simulations results.

Overall, the computed distribution of TKE are in a very good agreement with the
PIV data, considering the limitation of both the experimental setup and the
simulation (discussed below).

\subsection{Discussion}
\label{sec:disc}

Based on the current simulation results, the wall-channelling effect has a
strong influence over the whole flow, dictating the distribution of the
velocity both inside and over the bed. Above the bed, the simulated and
measured results are in good agreement. The differences
can be attributed to several effects:
\begin{enumerate}[i)]
  \item The modification of the geometry required for the meshing process could
    increase the velocity of the fluid flowing out of the packing by reducing
    the interstitial space. With that said, based on the comparisons of the
    interstitial flow, the chosen geometry simplification method seems to have a
    low impact on the accuracy of the results, but allow a more robust
    numerical setup. Since it is observed that the flow near the two topmost
    layers of spheres can have an influence on the conditions over the bed,
    there is, however, a chance that the simplification of geometry in this
    region can result in small discrepancies between the numerics and the
    simulation.
  \item The small inaccuracies in the dimensions of the experimental rig,
    resulting, among others, from the use of an  adhesive to stabilise the
    positions of the spheres, might lead to small differences between the
    numerical and experimental geometries as visible, for example, in
    Figure~\ref{fig:B2_L19_vel_mag}. The results measured by PIV have
    noticeable, but mostly inconsequential, imperfections, e.g.\, the breaking
    of symmetry of the measured velocity fields, which might be caused by the
    geometrical inaccuracies. We assume that such effects, which are also visible in the
    freeboard, can be perhaps attributed to these small shifts in the placement of the
    spheres.
  \item The mesh resolution chosen for the study strikes a balance between the
    required high accuracy and acceptable computation time to gather a sufficient
    amount of statistics describing the flow. The narrowest gap between the
    spheres in the topmost layer is discretised with around 10 cells (including
    the boundary layer mesh). It is possible that the resolution in these gaps
    in the top layers was not sufficient to fully resolve the gradients,
    leading to increased numerical diffusion and smearing of the ejected jets.
    A mesh resolution study, perhaps in a simplified setting, would be an
    important point of investigation for further studies of the described and
    similar configurations.
  \item The slight asymmetry of both experimental and simulated results can
    also be attributed to insufficiently converged mean velocity. The flow
    above the bed is dominated by low frequency oscillations and flapping
    motion~\citep{neeraj2023a}. The simulations were performed in such a way to
    enable the biggest possible statistical sample and match the averaging time of the experimental system. 
    However, observing the simulated TKE
    profile, it can be judged that more averaging time is certainly
    required for second-order statistics to converge.
\end{enumerate}

The simulated results allow for the inspection of the unsteady behaviour in areas
unavailable for the optical measurement techniques, and seem to indicate that
the onset of unsteady flow is located slightly below the top surface of the
packing. The spots of high TKE remain localised around the jets, potentially
due to the constrained geometry of the model packed-bed. A packing with
a bigger cross-section area w.r.t.\ the sphere size should be investigated to
determine if, in such a setup, the flow over the bed would become more
unstable. The experimental measurements and the computed results for both
\(\mathrm{Re}_p\) indicate that the flow near the walls remains mostly
steady up until \(y=13.5D_p\). The only exception to that is the centre plane
at \(\mathrm{Re}_p=300\), where the TKE near the walls is generated by the jets
being convected towards the walls. This result seems to indicate that the walls
have an overall stabilising effect on the unsteady flow field, which reinforces
the necessity for an analysis of a bigger packing.

The fact that the presented numerical setup results in a symmetric flow field
in the A1/20 region of interest, along with very well reproduced measurement
results in other interstices, indicate that the simulation is able to capture
the flow conditions inside the bed with remarkable accuracy, even if the
geometry of the packing is simplified. On the other hand, the fact that the 
results of the simulation are in agreement with the data measured in the packing 
with geometrical imperfections resulting from the assembly process reinforces 
that such an experimental approach is a robust and accurate method.

Overall, the flow inside the bed is dominated by the presence of recirculation
regions, whose shape and position are mainly influenced by the wall-channeled
fluid, which forces the flow into the interstices and in the direction of the
centre of the packing. Examining and understanding the behaviour of these
separated regions is important from the point of view of potential heat
transfer inside the bed, or more generally, mixing of species, as these phenomena
can lead to the presence of hotspots and uneven mixing processes. Although, the
current work does not attempt to perform such an in-depth analysis of scalar
mixing, the presented study of the flow inside the packing shows the
possibilities of investigating such flow properties when a PRS is performed
with conjunction with the experimental measurements.

\section{Conclusions}
\label{sec:conclusions}


In this study, Large Eddy Simulations were performed encompassing a packed bed
and its freeboard in the same configuration as investigated experimentally by
\citet{velten2023} using PIV, whereby a polyhedral, boundary
conforming mesh is employed to discretise the computational domain.

The simulated results compare very favourably with the data captured by PIV,
both inside and above the bed, however, small discrepancies between data sets
are visible. 
It can be argued that the reasons for this are twofold. On one hand, an exact
reproduction of the experimental geometry was not possible due to the measures
taken to stabilise the packing as described in Section
~\ref{sec:experimental-configuration}. On the other hand, the presence of
low-frequency flow structures developing in the freeboard and also observed in
\citealt{sadowski2023a,neeraj2023a} is deemed responsible for a very slow
convergence of the statistics.

Nevertheless, based on the gathered data, the particle-resolved simulation
conducted with the presented numerical setup has been seen to be a reliable tool for the
assessment of the flow conditions inside the packed bed. Combining the
CFD-generated prediction with the experimental measurements in the optically
accessible interstices allows for the validation of the CFD model and an in-depth
analysis of the flow field, as the areas unavailable to PIV can be potentially examined
using the validated simulation data. The simplification of the contact regions between the
individual spheres by means of the bridging method (described in
Section~\ref{sec:math-model}) allows for very accurate prediction of the
interstitial flow while sensibly reducing the problems related to mesh quality.

The flow investigated in the present work is, however, heavily dominated by  wall-channelling effects,
which steers both the flow inside the bed and in the freeboard, particularly determining the unsteady behaviour
over the packing. The jets of faster fluid ejected from the interstices near
the walls are the most important features of freeboard flow and generate
unsteady recirculation regions in the middle, above the packing. The onset of the
transient flow seems to be positioned below the crests of the spheres in the
top layer, indicating a complex behaviour of the fluid in the freeboard. To
further investigate the flow over the packed-beds, an experimental
configuration with a larger size-to-sphere diameter ratio should be investigated
both experimentally and numerically, as it would allow for the decrease of the
wall influence on the flow conditions.

\begin{acknowledgments}
  The authors gratefully acknowledge the financial support of the Deutsche
  Forschungsgemeinschaft (DFG) thorough SFB/TRR287, Project Number 422037413 
  The authors also gratefully acknowledge the
  Gauss Centre for Supercomputing e.V.\ (\url{https://www.gauss-centre.eu}) for providing computing time on the GCS Supercomputer SuperMUC--NG
  at the Leibniz Supercomputing Centre (\url{https://www.lrz.de}).
\end{acknowledgments}

\section*{Data Availability Statement}
The data that support the findings of this study are available from the
corresponding author upon reasonable request.

\appendix

\section*{References}
\bibliography{ms}

\begin{thebibliography}{49}%
\makeatletter
\providecommand \@ifxundefined [1]{%
 \@ifx{#1\undefined}
}%
\providecommand \@ifnum [1]{%
 \ifnum #1\expandafter \@firstoftwo
 \else \expandafter \@secondoftwo
 \fi
}%
\providecommand \@ifx [1]{%
 \ifx #1\expandafter \@firstoftwo
 \else \expandafter \@secondoftwo
 \fi
}%
\providecommand \natexlab [1]{#1}%
\providecommand \enquote  [1]{``#1''}%
\providecommand \bibnamefont  [1]{#1}%
\providecommand \bibfnamefont [1]{#1}%
\providecommand \citenamefont [1]{#1}%
\providecommand \href@noop [0]{\@secondoftwo}%
\providecommand \href [0]{\begingroup \@sanitize@url \@href}%
\providecommand \@href[1]{\@@startlink{#1}\@@href}%
\providecommand \@@href[1]{\endgroup#1\@@endlink}%
\providecommand \@sanitize@url [0]{\catcode `\\12\catcode `\$12\catcode
  `\&12\catcode `\#12\catcode `\^12\catcode `\_12\catcode `\%12\relax}%
\providecommand \@@startlink[1]{}%
\providecommand \@@endlink[0]{}%
\providecommand \url  [0]{\begingroup\@sanitize@url \@url }%
\providecommand \@url [1]{\endgroup\@href {#1}{\urlprefix }}%
\providecommand \urlprefix  [0]{URL }%
\providecommand \Eprint [0]{\href }%
\providecommand \doibase [0]{http://dx.doi.org/}%
\providecommand \selectlanguage [0]{\@gobble}%
\providecommand \bibinfo  [0]{\@secondoftwo}%
\providecommand \bibfield  [0]{\@secondoftwo}%
\providecommand \translation [1]{[#1]}%
\providecommand \BibitemOpen [0]{}%
\providecommand \bibitemStop [0]{}%
\providecommand \bibitemNoStop [0]{.\EOS\space}%
\providecommand \EOS [0]{\spacefactor3000\relax}%
\providecommand \BibitemShut  [1]{\csname bibitem#1\endcsname}%
\let\auto@bib@innerbib\@empty
\bibitem [{\citenamefont {Velten}\ and\ \citenamefont
  {Z{\"a}hringer}(2023)}]{velten2023}%
  \BibitemOpen
  \bibfield  {author} {\bibinfo {author} {\bibfnamefont {C.}~\bibnamefont
  {Velten}}\ and\ \bibinfo {author} {\bibfnamefont {K.}~\bibnamefont
  {Z{\"a}hringer}},\ }\bibfield  {title} {\enquote {\bibinfo {title} {Flow
  {{Field Characterisation}} of {{Gaseous Flow}} in a {{Packed Bed}} by
  {{Particle Image Velocimetry}}},}\ }\href {\doibase
  10.1007/s11242-023-02010-7} {\bibfield  {journal} {\bibinfo  {journal}
  {Transport in Porous Media}\ } (\bibinfo {year} {2023}),\
  10.1007/s11242-023-02010-7}\BibitemShut {NoStop}%
\bibitem [{\citenamefont {Shiehnejadhesar}\ \emph {et~al.}(2014)\citenamefont
  {Shiehnejadhesar}, \citenamefont {Mehrabian}, \citenamefont {Scharler},
  \citenamefont {Goldin},\ and\ \citenamefont
  {Obernberger}}]{shiehnejadhesar2014}%
  \BibitemOpen
  \bibfield  {author} {\bibinfo {author} {\bibfnamefont {A.}~\bibnamefont
  {Shiehnejadhesar}}, \bibinfo {author} {\bibfnamefont {R.}~\bibnamefont
  {Mehrabian}}, \bibinfo {author} {\bibfnamefont {R.}~\bibnamefont {Scharler}},
  \bibinfo {author} {\bibfnamefont {G.~M.}\ \bibnamefont {Goldin}}, \ and\
  \bibinfo {author} {\bibfnamefont {I.}~\bibnamefont {Obernberger}},\
  }\bibfield  {title} {\enquote {\bibinfo {title} {Development of a gas phase
  combustion model suitable for low and high turbulence conditions},}\ }\href
  {\doibase 10.1016/j.fuel.2014.02.040} {\bibfield  {journal} {\bibinfo
  {journal} {Fuel}\ }\textbf {\bibinfo {volume} {126}},\ \bibinfo {pages}
  {177--187} (\bibinfo {year} {2014})}\BibitemShut {NoStop}%
\bibitem [{\citenamefont {Dixon}\ and\ \citenamefont
  {Partopour}(2020)}]{dixon2020}%
  \BibitemOpen
  \bibfield  {author} {\bibinfo {author} {\bibfnamefont {A.~G.}\ \bibnamefont
  {Dixon}}\ and\ \bibinfo {author} {\bibfnamefont {B.}~\bibnamefont
  {Partopour}},\ }\bibfield  {title} {\enquote {\bibinfo {title} {Computational
  {Fluid Dynamics} for {Fixed Bed Reactor Design}},}\ }\href {\doibase
  10.1146/annurev-chembioeng-092319-075328} {\bibfield  {journal} {\bibinfo
  {journal} {Annual Review of Chemical and Biomolecular Engineering}\ }\textbf
  {\bibinfo {volume} {11}},\ \bibinfo {pages} {109--130} (\bibinfo {year}
  {2020})}\BibitemShut {NoStop}%
\bibitem [{\citenamefont {Dixon}, \citenamefont {Nijemeisland},\ and\
  \citenamefont {Stitt}(2006)}]{dixon2006}%
  \BibitemOpen
  \bibfield  {author} {\bibinfo {author} {\bibfnamefont {A.~G.}\ \bibnamefont
  {Dixon}}, \bibinfo {author} {\bibfnamefont {M.}~\bibnamefont {Nijemeisland}},
  \ and\ \bibinfo {author} {\bibfnamefont {E.~H.}\ \bibnamefont {Stitt}},\
  }\bibfield  {title} {\enquote {\bibinfo {title} {Packed {{Tubular Reactor
  Modeling}} and {{Catalyst Design}} using {{Computational Fluid Dynamics}}},}\
  }in\ \href {\doibase 10.1016/S0065-2377(06)31005-8} {\emph {\bibinfo
  {booktitle} {Advances in {{Chemical Engineering}}}}},\ Vol.~\bibinfo {volume}
  {31}\ (\bibinfo  {publisher} {{Elsevier}},\ \bibinfo {year} {2006})\ pp.\
  \bibinfo {pages} {307--389}\BibitemShut {NoStop}%
\bibitem [{\citenamefont {Jurtz}, \citenamefont {Kraume},\ and\ \citenamefont
  {Wehinger}(2019)}]{jurtz2019}%
  \BibitemOpen
  \bibfield  {author} {\bibinfo {author} {\bibfnamefont {N.}~\bibnamefont
  {Jurtz}}, \bibinfo {author} {\bibfnamefont {M.}~\bibnamefont {Kraume}}, \
  and\ \bibinfo {author} {\bibfnamefont {G.~D.}\ \bibnamefont {Wehinger}},\
  }\bibfield  {title} {\enquote {\bibinfo {title} {Advances in fixed-bed
  reactor modeling using particle-resolved computational fluid dynamics
  ({{CFD}})},}\ }\href {\doibase 10.1515/revce-2017-0059} {\bibfield  {journal}
  {\bibinfo  {journal} {Reviews in Chemical Engineering}\ }\textbf {\bibinfo
  {volume} {35}},\ \bibinfo {pages} {139--190} (\bibinfo {year}
  {2019})}\BibitemShut {NoStop}%
\bibitem [{\citenamefont {Shams}\ \emph
  {et~al.}(2013{\natexlab{a}})\citenamefont {Shams}, \citenamefont {Roelofs},
  \citenamefont {Komen},\ and\ \citenamefont {Baglietto}}]{shams2013}%
  \BibitemOpen
  \bibfield  {author} {\bibinfo {author} {\bibfnamefont {A.}~\bibnamefont
  {Shams}}, \bibinfo {author} {\bibfnamefont {F.}~\bibnamefont {Roelofs}},
  \bibinfo {author} {\bibfnamefont {E.}~\bibnamefont {Komen}}, \ and\ \bibinfo
  {author} {\bibfnamefont {E.}~\bibnamefont {Baglietto}},\ }\bibfield  {title}
  {\enquote {\bibinfo {title} {Large eddy simulation of a nuclear pebble bed
  configuration},}\ }\href {\doibase 10.1016/j.nucengdes.2013.03.040}
  {\bibfield  {journal} {\bibinfo  {journal} {Nuclear Engineering and Design}\
  }\textbf {\bibinfo {volume} {261}},\ \bibinfo {pages} {10--19} (\bibinfo
  {year} {2013}{\natexlab{a}})}\BibitemShut {NoStop}%
\bibitem [{\citenamefont {Shams}\ \emph
  {et~al.}(2013{\natexlab{b}})\citenamefont {Shams}, \citenamefont {Roelofs},
  \citenamefont {Komen},\ and\ \citenamefont {Baglietto}}]{shams2013a}%
  \BibitemOpen
  \bibfield  {author} {\bibinfo {author} {\bibfnamefont {A.}~\bibnamefont
  {Shams}}, \bibinfo {author} {\bibfnamefont {F.}~\bibnamefont {Roelofs}},
  \bibinfo {author} {\bibfnamefont {E.}~\bibnamefont {Komen}}, \ and\ \bibinfo
  {author} {\bibfnamefont {E.}~\bibnamefont {Baglietto}},\ }\bibfield  {title}
  {\enquote {\bibinfo {title} {Numerical simulations of a pebble bed
  configuration using hybrid ({{RANS}}\textendash{{LES}}) methods},}\ }\href
  {\doibase 10.1016/j.nucengdes.2013.04.001} {\bibfield  {journal} {\bibinfo
  {journal} {Nuclear Engineering and Design}\ }\textbf {\bibinfo {volume}
  {261}},\ \bibinfo {pages} {201--211} (\bibinfo {year}
  {2013}{\natexlab{b}})}\BibitemShut {NoStop}%
\bibitem [{\citenamefont {Shams}\ \emph
  {et~al.}(2013{\natexlab{c}})\citenamefont {Shams}, \citenamefont {Roelofs},
  \citenamefont {Komen},\ and\ \citenamefont {Baglietto}}]{shams2013b}%
  \BibitemOpen
  \bibfield  {author} {\bibinfo {author} {\bibfnamefont {A.}~\bibnamefont
  {Shams}}, \bibinfo {author} {\bibfnamefont {F.}~\bibnamefont {Roelofs}},
  \bibinfo {author} {\bibfnamefont {E.}~\bibnamefont {Komen}}, \ and\ \bibinfo
  {author} {\bibfnamefont {E.}~\bibnamefont {Baglietto}},\ }\bibfield  {title}
  {\enquote {\bibinfo {title} {Quasi-direct numerical simulation of a pebble
  bed configuration, {{Part-II}}: {{Temperature}} field analysis},}\ }\href
  {\doibase 10.1016/j.nucengdes.2013.02.015} {\bibfield  {journal} {\bibinfo
  {journal} {Nuclear Engineering and Design}\ }\textbf {\bibinfo {volume}
  {263}},\ \bibinfo {pages} {490--499} (\bibinfo {year}
  {2013}{\natexlab{c}})}\BibitemShut {NoStop}%
\bibitem [{\citenamefont {Shams}\ \emph
  {et~al.}(2013{\natexlab{d}})\citenamefont {Shams}, \citenamefont {Roelofs},
  \citenamefont {Komen},\ and\ \citenamefont {Baglietto}}]{shams2013c}%
  \BibitemOpen
  \bibfield  {author} {\bibinfo {author} {\bibfnamefont {A.}~\bibnamefont
  {Shams}}, \bibinfo {author} {\bibfnamefont {F.}~\bibnamefont {Roelofs}},
  \bibinfo {author} {\bibfnamefont {E.}~\bibnamefont {Komen}}, \ and\ \bibinfo
  {author} {\bibfnamefont {E.}~\bibnamefont {Baglietto}},\ }\bibfield  {title}
  {\enquote {\bibinfo {title} {Quasi-direct numerical simulation of a pebble
  bed configuration. {{Part I}}: {{Flow}} (velocity) field analysis},}\ }\href
  {\doibase 10.1016/j.nucengdes.2012.06.016} {\bibfield  {journal} {\bibinfo
  {journal} {Nuclear Engineering and Design}\ }\textbf {\bibinfo {volume}
  {263}},\ \bibinfo {pages} {473--489} (\bibinfo {year}
  {2013}{\natexlab{d}})}\BibitemShut {NoStop}%
\bibitem [{\citenamefont {Shams}\ \emph {et~al.}(2015)\citenamefont {Shams},
  \citenamefont {Roelofs}, \citenamefont {Komen},\ and\ \citenamefont
  {Baglietto}}]{shams2015}%
  \BibitemOpen
  \bibfield  {author} {\bibinfo {author} {\bibfnamefont {A.}~\bibnamefont
  {Shams}}, \bibinfo {author} {\bibfnamefont {F.}~\bibnamefont {Roelofs}},
  \bibinfo {author} {\bibfnamefont {E.}~\bibnamefont {Komen}}, \ and\ \bibinfo
  {author} {\bibfnamefont {E.}~\bibnamefont {Baglietto}},\ }\bibfield  {title}
  {\enquote {\bibinfo {title} {Improved delayed detached eddy simulation of a
  randomly stacked nuclear pebble bed},}\ }\href {\doibase
  10.1016/j.compfluid.2015.08.015} {\bibfield  {journal} {\bibinfo  {journal}
  {Computers \& Fluids}\ }\textbf {\bibinfo {volume} {122}},\ \bibinfo {pages}
  {12--25} (\bibinfo {year} {2015})}\BibitemShut {NoStop}%
\bibitem [{\citenamefont {Jin}\ \emph {et~al.}(2015)\citenamefont {Jin},
  \citenamefont {Uth}, \citenamefont {Kuznetsov},\ and\ \citenamefont
  {Herwig}}]{jin2015}%
  \BibitemOpen
  \bibfield  {author} {\bibinfo {author} {\bibfnamefont {Y.}~\bibnamefont
  {Jin}}, \bibinfo {author} {\bibfnamefont {M.-F.}\ \bibnamefont {Uth}},
  \bibinfo {author} {\bibfnamefont {A.~V.}\ \bibnamefont {Kuznetsov}}, \ and\
  \bibinfo {author} {\bibfnamefont {H.}~\bibnamefont {Herwig}},\ }\bibfield
  {title} {\enquote {\bibinfo {title} {Numerical investigation of the
  possibility of macroscopic turbulence in porous media: a direct numerical
  simulation study},}\ }\href {\doibase 10.1017/jfm.2015.9} {\bibfield
  {journal} {\bibinfo  {journal} {Journal of Fluid Mechanics}\ }\textbf
  {\bibinfo {volume} {766}},\ \bibinfo {pages} {76--103} (\bibinfo {year}
  {2015})}\BibitemShut {NoStop}%
\bibitem [{\citenamefont {Uth}\ \emph {et~al.}(2016)\citenamefont {Uth},
  \citenamefont {Jin}, \citenamefont {Kuznetsov},\ and\ \citenamefont
  {Herwig}}]{uth2016}%
  \BibitemOpen
  \bibfield  {author} {\bibinfo {author} {\bibfnamefont {M.-F.}\ \bibnamefont
  {Uth}}, \bibinfo {author} {\bibfnamefont {Y.}~\bibnamefont {Jin}}, \bibinfo
  {author} {\bibfnamefont {A.~V.}\ \bibnamefont {Kuznetsov}}, \ and\ \bibinfo
  {author} {\bibfnamefont {H.}~\bibnamefont {Herwig}},\ }\bibfield  {title}
  {\enquote {\bibinfo {title} {A direct numerical simulation study on the
  possibility of macroscopic turbulence in porous media: {Effects} of different
  solid matrix geometries, solid boundaries, and two porosity scales},}\ }\href
  {\doibase 10.1063/1.4949549} {\bibfield  {journal} {\bibinfo  {journal}
  {Physics of Fluids}\ }\textbf {\bibinfo {volume} {28}} (\bibinfo {year}
  {2016}),\ 10.1063/1.4949549}\BibitemShut {NoStop}%
\bibitem [{\citenamefont {Rao}\ and\ \citenamefont {Jin}(2022)}]{rao2022}%
  \BibitemOpen
  \bibfield  {author} {\bibinfo {author} {\bibfnamefont {F.}~\bibnamefont
  {Rao}}\ and\ \bibinfo {author} {\bibfnamefont {Y.}~\bibnamefont {Jin}},\
  }\bibfield  {title} {\enquote {\bibinfo {title} {Possibility for survival of
  macroscopic turbulence in porous media with high porosity},}\ }\href
  {\doibase 10.1017/jfm.2022.87} {\bibfield  {journal} {\bibinfo  {journal}
  {Journal of Fluid Mechanics}\ }\textbf {\bibinfo {volume} {937}} (\bibinfo
  {year} {2022}),\ 10.1017/jfm.2022.87}\BibitemShut {NoStop}%
\bibitem [{\citenamefont {Lage}, \citenamefont {De~Lemos},\ and\ \citenamefont
  {Nield}(2002)}]{lage2002}%
  \BibitemOpen
  \bibfield  {author} {\bibinfo {author} {\bibfnamefont {J.}~\bibnamefont
  {Lage}}, \bibinfo {author} {\bibfnamefont {M.}~\bibnamefont {De~Lemos}}, \
  and\ \bibinfo {author} {\bibfnamefont {D.}~\bibnamefont {Nield}},\ }\bibfield
   {title} {\enquote {\bibinfo {title} {Modeling {{Turbulence}} in {{Porous
  Media}}},}\ }in\ \href {\doibase 10.1016/B978-008043965-5/50009-X} {\emph
  {\bibinfo {booktitle} {Transport {{Phenomena}} in {{Porous Media II}}}}}\
  (\bibinfo  {publisher} {{Elsevier}},\ \bibinfo {year} {2002})\ pp.\ \bibinfo
  {pages} {198--230}\BibitemShut {NoStop}%
\bibitem [{\citenamefont {Sadowski}\ \emph {et~al.}(2023)\citenamefont
  {Sadowski}, \citenamefont {Sayyari}, \citenamefont {Di~Mare},\ and\
  \citenamefont {Marschall}}]{sadowski2023a}%
  \BibitemOpen
  \bibfield  {author} {\bibinfo {author} {\bibfnamefont {W.}~\bibnamefont
  {Sadowski}}, \bibinfo {author} {\bibfnamefont {M.}~\bibnamefont {Sayyari}},
  \bibinfo {author} {\bibfnamefont {F.}~\bibnamefont {Di~Mare}}, \ and\
  \bibinfo {author} {\bibfnamefont {H.}~\bibnamefont {Marschall}},\ }\bibfield
  {title} {\enquote {\bibinfo {title} {Large eddy simulation of flow in porous
  media: {{Analysis}} of the commutation error of the double-averaged
  equations},}\ }\href {\doibase 10.1063/5.0148130} {\bibfield  {journal}
  {\bibinfo  {journal} {Physics of Fluids}\ }\textbf {\bibinfo {volume} {35}},\
  \bibinfo {pages} {055121} (\bibinfo {year} {2023})}\BibitemShut {NoStop}%
\bibitem [{\citenamefont {Breugem}\ and\ \citenamefont
  {Boersma}(2005)}]{breugem2005}%
  \BibitemOpen
  \bibfield  {author} {\bibinfo {author} {\bibfnamefont {W.~P.}\ \bibnamefont
  {Breugem}}\ and\ \bibinfo {author} {\bibfnamefont {B.~J.}\ \bibnamefont
  {Boersma}},\ }\bibfield  {title} {\enquote {\bibinfo {title} {Direct
  numerical simulations of turbulent flow over a permeable wall using a direct
  and a continuum approach},}\ }\href {\doibase 10.1063/1.1835771} {\bibfield
  {journal} {\bibinfo  {journal} {Physics of Fluids}\ }\textbf {\bibinfo
  {volume} {17}},\ \bibinfo {pages} {025103} (\bibinfo {year}
  {2005})}\BibitemShut {NoStop}%
\bibitem [{\citenamefont {Kuwata}\ and\ \citenamefont
  {Suga}(2016)}]{kuwata2016}%
  \BibitemOpen
  \bibfield  {author} {\bibinfo {author} {\bibfnamefont {Y.}~\bibnamefont
  {Kuwata}}\ and\ \bibinfo {author} {\bibfnamefont {K.}~\bibnamefont {Suga}},\
  }\bibfield  {title} {\enquote {\bibinfo {title} {Lattice {{Boltzmann}} direct
  numerical simulation of interface turbulence over porous and rough walls},}\
  }\href {\doibase 10.1016/j.ijheatfluidflow.2016.03.006} {\bibfield  {journal}
  {\bibinfo  {journal} {International Journal of Heat and Fluid Flow}\ }\textbf
  {\bibinfo {volume} {61}},\ \bibinfo {pages} {145--157} (\bibinfo {year}
  {2016})}\BibitemShut {NoStop}%
\bibitem [{\citenamefont {Shams}\ \emph {et~al.}(2014)\citenamefont {Shams},
  \citenamefont {Roelofs}, \citenamefont {Komen},\ and\ \citenamefont
  {Baglietto}}]{shams2014_stacked}%
  \BibitemOpen
  \bibfield  {author} {\bibinfo {author} {\bibfnamefont {A.}~\bibnamefont
  {Shams}}, \bibinfo {author} {\bibfnamefont {F.}~\bibnamefont {Roelofs}},
  \bibinfo {author} {\bibfnamefont {E.}~\bibnamefont {Komen}}, \ and\ \bibinfo
  {author} {\bibfnamefont {E.}~\bibnamefont {Baglietto}},\ }\bibfield  {title}
  {\enquote {\bibinfo {title} {Large eddy simulation of a randomly stacked
  nuclear pebble bed},}\ }\href {\doibase 10.1016/j.compfluid.2014.03.025}
  {\bibfield  {journal} {\bibinfo  {journal} {Computers \& Fluids}\ }\textbf
  {\bibinfo {volume} {96}},\ \bibinfo {pages} {302--321} (\bibinfo {year}
  {2014})}\BibitemShut {NoStop}%
\bibitem [{\citenamefont {Rebughini}, \citenamefont {Cuoci},\ and\
  \citenamefont {Maestri}(2016)}]{rebughini2016}%
  \BibitemOpen
  \bibfield  {author} {\bibinfo {author} {\bibfnamefont {S.}~\bibnamefont
  {Rebughini}}, \bibinfo {author} {\bibfnamefont {A.}~\bibnamefont {Cuoci}}, \
  and\ \bibinfo {author} {\bibfnamefont {M.}~\bibnamefont {Maestri}},\
  }\bibfield  {title} {\enquote {\bibinfo {title} {Handling contact points in
  reactive cfd simulations of heterogeneous catalytic fixed bed reactors},}\
  }\href {\doibase https://doi.org/10.1016/j.ces.2015.11.013} {\bibfield
  {journal} {\bibinfo  {journal} {Chemical Engineering Science}\ }\textbf
  {\bibinfo {volume} {141}},\ \bibinfo {pages} {240--249} (\bibinfo {year}
  {2016})}\BibitemShut {NoStop}%
\bibitem [{\citenamefont {Gorges}\ \emph {et~al.}(2024)\citenamefont {Gorges},
  \citenamefont {Br{\"o}mmer}, \citenamefont {Velten}, \citenamefont {Wirtz},
  \citenamefont {Mahiques}, \citenamefont {Scherer}, \citenamefont
  {Z{\"a}hringer},\ and\ \citenamefont {van Wachem}}]{gorges2024comparing}%
  \BibitemOpen
  \bibfield  {author} {\bibinfo {author} {\bibfnamefont {C.}~\bibnamefont
  {Gorges}}, \bibinfo {author} {\bibfnamefont {M.}~\bibnamefont {Br{\"o}mmer}},
  \bibinfo {author} {\bibfnamefont {C.}~\bibnamefont {Velten}}, \bibinfo
  {author} {\bibfnamefont {S.}~\bibnamefont {Wirtz}}, \bibinfo {author}
  {\bibfnamefont {E.~I.}\ \bibnamefont {Mahiques}}, \bibinfo {author}
  {\bibfnamefont {V.}~\bibnamefont {Scherer}}, \bibinfo {author} {\bibfnamefont
  {K.}~\bibnamefont {Z{\"a}hringer}}, \ and\ \bibinfo {author} {\bibfnamefont
  {B.}~\bibnamefont {van Wachem}},\ }\bibfield  {title} {\enquote {\bibinfo
  {title} {Comparing two ibm implementations for the simulation of uniform
  packed beds},}\ }\href@noop {} {\bibfield  {journal} {\bibinfo  {journal}
  {Particuology}\ }\textbf {\bibinfo {volume} {86}},\ \bibinfo {pages} {1--12}
  (\bibinfo {year} {2024})}\BibitemShut {NoStop}%
\bibitem [{\citenamefont {Neeraj}\ \emph {et~al.}(2023)\citenamefont {Neeraj},
  \citenamefont {Velten}, \citenamefont {Janiga}, \citenamefont
  {Z{\"a}hringer}, \citenamefont {Namdar}, \citenamefont {Varnik},
  \citenamefont {Th{\'e}venin},\ and\ \citenamefont {Hosseini}}]{neeraj2023a}%
  \BibitemOpen
  \bibfield  {author} {\bibinfo {author} {\bibfnamefont {T.}~\bibnamefont
  {Neeraj}}, \bibinfo {author} {\bibfnamefont {C.}~\bibnamefont {Velten}},
  \bibinfo {author} {\bibfnamefont {G.}~\bibnamefont {Janiga}}, \bibinfo
  {author} {\bibfnamefont {K.}~\bibnamefont {Z{\"a}hringer}}, \bibinfo {author}
  {\bibfnamefont {R.}~\bibnamefont {Namdar}}, \bibinfo {author} {\bibfnamefont
  {F.}~\bibnamefont {Varnik}}, \bibinfo {author} {\bibfnamefont
  {D.}~\bibnamefont {Th{\'e}venin}}, \ and\ \bibinfo {author} {\bibfnamefont
  {S.~A.}\ \bibnamefont {Hosseini}},\ }\bibfield  {title} {\enquote {\bibinfo
  {title} {Modeling {{Gas Flows}} in {{Packed Beds}} with the {{Lattice
  Boltzmann Method}}: {{Validation Against Experiments}}},}\ }\href {\doibase
  10.1007/s10494-023-00444-z} {\bibfield  {journal} {\bibinfo  {journal} {Flow,
  Turbulence and Combustion}\ }\textbf {\bibinfo {volume} {111}},\ \bibinfo
  {pages} {463--491} (\bibinfo {year} {2023})}\BibitemShut {NoStop}%
\bibitem [{\citenamefont {Wongkham}\ \emph {et~al.}(2020)\citenamefont
  {Wongkham}, \citenamefont {Wen}, \citenamefont {Lu}, \citenamefont {Cui},
  \citenamefont {Xu},\ and\ \citenamefont {Liu}}]{wongkham2020}%
  \BibitemOpen
  \bibfield  {author} {\bibinfo {author} {\bibfnamefont {J.}~\bibnamefont
  {Wongkham}}, \bibinfo {author} {\bibfnamefont {T.}~\bibnamefont {Wen}},
  \bibinfo {author} {\bibfnamefont {B.}~\bibnamefont {Lu}}, \bibinfo {author}
  {\bibfnamefont {L.}~\bibnamefont {Cui}}, \bibinfo {author} {\bibfnamefont
  {J.}~\bibnamefont {Xu}}, \ and\ \bibinfo {author} {\bibfnamefont
  {X.}~\bibnamefont {Liu}},\ }\bibfield  {title} {\enquote {\bibinfo {title}
  {Particle-resolved simulation of randomly packed pebble beds with a novel
  fluid-solid coupling method},}\ }\href {\doibase
  10.1016/j.fusengdes.2020.111953} {\bibfield  {journal} {\bibinfo  {journal}
  {Fusion Engineering and Design}\ }\textbf {\bibinfo {volume} {161}},\
  \bibinfo {pages} {111953} (\bibinfo {year} {2020})}\BibitemShut {NoStop}%
\bibitem [{\citenamefont {Giese}, \citenamefont {Rottsch\"{a}fer},\ and\
  \citenamefont {Vortmeyer}(1998)}]{giese1998}%
  \BibitemOpen
  \bibfield  {author} {\bibinfo {author} {\bibfnamefont {M.}~\bibnamefont
  {Giese}}, \bibinfo {author} {\bibfnamefont {K.}~\bibnamefont
  {Rottsch\"{a}fer}}, \ and\ \bibinfo {author} {\bibfnamefont {D.}~\bibnamefont
  {Vortmeyer}},\ }\bibfield  {title} {\enquote {\bibinfo {title} {Measured and
  modeled superficial flow profiles in packed beds with liquid flow},}\ }\href
  {\doibase 10.1002/aic.690440225} {\bibfield  {journal} {\bibinfo  {journal}
  {AIChE Journal}\ }\textbf {\bibinfo {volume} {44}},\ \bibinfo {pages}
  {484--490} (\bibinfo {year} {1998})}\BibitemShut {NoStop}%
\bibitem [{\citenamefont {Wehinger}(2016)}]{wehinger2016}%
  \BibitemOpen
  \bibfield  {author} {\bibinfo {author} {\bibfnamefont {G.~D.}\ \bibnamefont
  {Wehinger}},\ }\emph {\bibinfo {title} {Particle-resolved {CFD} simulations
  of catalytic flow reactors}},\ \href@noop {} {\bibinfo {type} {Doctoral}},\
  \bibinfo  {school} {TU Berlin} (\bibinfo {year} {2016})\BibitemShut {NoStop}%
\bibitem [{\citenamefont {Morales}, \citenamefont {Spinn},\ and\ \citenamefont
  {Smith}(1951)}]{Morales1951}%
  \BibitemOpen
  \bibfield  {author} {\bibinfo {author} {\bibfnamefont {M.}~\bibnamefont
  {Morales}}, \bibinfo {author} {\bibfnamefont {C.~W.}\ \bibnamefont {Spinn}},
  \ and\ \bibinfo {author} {\bibfnamefont {J.~M.}\ \bibnamefont {Smith}},\
  }\bibfield  {title} {\enquote {\bibinfo {title} {Velocities and effective
  thermal conductivities in packed beds},}\ }\href {\doibase
  10.1021/ie50493a055} {\bibfield  {journal} {\bibinfo  {journal} {Industrial
  {\&} Engineering Chemistry}\ }\textbf {\bibinfo {volume} {43}},\ \bibinfo
  {pages} {225--232} (\bibinfo {year} {1951})}\BibitemShut {NoStop}%
\bibitem [{\citenamefont {Schwartz}\ and\ \citenamefont
  {Smith}(1953)}]{Schwartz1953}%
  \BibitemOpen
  \bibfield  {author} {\bibinfo {author} {\bibfnamefont {C.~E.}\ \bibnamefont
  {Schwartz}}\ and\ \bibinfo {author} {\bibfnamefont {J.~M.}\ \bibnamefont
  {Smith}},\ }\bibfield  {title} {\enquote {\bibinfo {title} {Flow distribution
  in packed beds},}\ }\href@noop {} {\bibfield  {journal} {\bibinfo  {journal}
  {Industrial {\&} Engineering Chemistry}\ }\textbf {\bibinfo {volume} {45}},\
  \bibinfo {pages} {1209--1218} (\bibinfo {year} {1953})}\BibitemShut {NoStop}%
\bibitem [{\citenamefont {Blois}\ \emph {et~al.}(2012)\citenamefont {Blois},
  \citenamefont {{Sambrook Smith}}, \citenamefont {Best}, \citenamefont
  {Hardy},\ and\ \citenamefont {Lead}}]{Blois2012}%
  \BibitemOpen
  \bibfield  {author} {\bibinfo {author} {\bibfnamefont {G.}~\bibnamefont
  {Blois}}, \bibinfo {author} {\bibfnamefont {G.~H.}\ \bibnamefont {{Sambrook
  Smith}}}, \bibinfo {author} {\bibfnamefont {J.~L.}\ \bibnamefont {Best}},
  \bibinfo {author} {\bibfnamefont {R.~J.}\ \bibnamefont {Hardy}}, \ and\
  \bibinfo {author} {\bibfnamefont {J.~R.}\ \bibnamefont {Lead}},\ }\bibfield
  {title} {\enquote {\bibinfo {title} {Quantifying the dynamics of flow within
  a permeable bed using time-resolved endoscopic particle imaging velocimetry
  (epiv)},}\ }\href {\doibase 10.1007/s00348-011-1198-8} {\bibfield  {journal}
  {\bibinfo  {journal} {Experiments in Fluids}\ }\textbf {\bibinfo {volume}
  {53}},\ \bibinfo {pages} {51--76} (\bibinfo {year} {2012})}\BibitemShut
  {NoStop}%
\bibitem [{\citenamefont {Sederman}\ \emph {et~al.}(1997)\citenamefont
  {Sederman}, \citenamefont {Johns}, \citenamefont {Bramley}, \citenamefont
  {Alexander},\ and\ \citenamefont {Gladden}}]{Sederman1997}%
  \BibitemOpen
  \bibfield  {author} {\bibinfo {author} {\bibfnamefont {A.~J.}\ \bibnamefont
  {Sederman}}, \bibinfo {author} {\bibfnamefont {M.~L.}\ \bibnamefont {Johns}},
  \bibinfo {author} {\bibfnamefont {A.~S.}\ \bibnamefont {Bramley}}, \bibinfo
  {author} {\bibfnamefont {P.}~\bibnamefont {Alexander}}, \ and\ \bibinfo
  {author} {\bibfnamefont {L.~F.}\ \bibnamefont {Gladden}},\ }\bibfield
  {title} {\enquote {\bibinfo {title} {Magnetic resonance imaging of liquid
  flow and pore structure within packed beds},}\ }\href {\doibase
  10.1016/S0009-2509(97)00057-2} {\bibfield  {journal} {\bibinfo  {journal}
  {Chemical Engineering Science}\ }\textbf {\bibinfo {volume} {52}},\ \bibinfo
  {pages} {2239--2250} (\bibinfo {year} {1997})}\BibitemShut {NoStop}%
\bibitem [{\citenamefont {Suekane}, \citenamefont {Yokouchi},\ and\
  \citenamefont {Hirai}(2003)}]{Suekane2003}%
  \BibitemOpen
  \bibfield  {author} {\bibinfo {author} {\bibfnamefont {T.}~\bibnamefont
  {Suekane}}, \bibinfo {author} {\bibfnamefont {Y.}~\bibnamefont {Yokouchi}}, \
  and\ \bibinfo {author} {\bibfnamefont {S.}~\bibnamefont {Hirai}},\ }\bibfield
   {title} {\enquote {\bibinfo {title} {Inertial flow structures in a
  simple-packed bed of spheres},}\ }\href {\doibase 10.1002/aic.690490103}
  {\bibfield  {journal} {\bibinfo  {journal} {AIChE Journal}\ }\textbf
  {\bibinfo {volume} {49}},\ \bibinfo {pages} {10--17} (\bibinfo {year}
  {2003})}\BibitemShut {NoStop}%
\bibitem [{\citenamefont {Lovreglio}\ \emph {et~al.}(2018)\citenamefont
  {Lovreglio}, \citenamefont {Das}, \citenamefont {Buist}, \citenamefont
  {Peters}, \citenamefont {Pel},\ and\ \citenamefont
  {Kuipers}}]{Lovreglio2018}%
  \BibitemOpen
  \bibfield  {author} {\bibinfo {author} {\bibfnamefont {P.}~\bibnamefont
  {Lovreglio}}, \bibinfo {author} {\bibfnamefont {S.}~\bibnamefont {Das}},
  \bibinfo {author} {\bibfnamefont {K.~A.}\ \bibnamefont {Buist}}, \bibinfo
  {author} {\bibfnamefont {E.~A. J.~F.}\ \bibnamefont {Peters}}, \bibinfo
  {author} {\bibfnamefont {L.}~\bibnamefont {Pel}}, \ and\ \bibinfo {author}
  {\bibfnamefont {J.~A.~M.}\ \bibnamefont {Kuipers}},\ }\bibfield  {title}
  {\enquote {\bibinfo {title} {Experimental and numerical investigation of
  structure and hydrodynamics in packed beds of spherical particles},}\ }\href
  {\doibase 10.1002/aic.16127} {\bibfield  {journal} {\bibinfo  {journal}
  {AIChE Journal}\ }\textbf {\bibinfo {volume} {64}},\ \bibinfo {pages}
  {1896--1907} (\bibinfo {year} {2018})}\BibitemShut {NoStop}%
\bibitem [{\citenamefont {Pokrajac}\ and\ \citenamefont
  {Manes}(2009)}]{Pokrajac2009}%
  \BibitemOpen
  \bibfield  {author} {\bibinfo {author} {\bibfnamefont {D.}~\bibnamefont
  {Pokrajac}}\ and\ \bibinfo {author} {\bibfnamefont {C.}~\bibnamefont
  {Manes}},\ }\bibfield  {title} {\enquote {\bibinfo {title} {Velocity
  measurements of a free-surface turbulent flow penetrating a porous medium
  composed of uniform-size spheres},}\ }\href {\doibase
  10.1007/s11242-009-9339-8} {\bibfield  {journal} {\bibinfo  {journal}
  {Transport in Porous Media}\ }\textbf {\bibinfo {volume} {78}},\ \bibinfo
  {pages} {367--383} (\bibinfo {year} {2009})}\BibitemShut {NoStop}%
\bibitem [{\citenamefont {Khayamyan}\ \emph
  {et~al.}(2017{\natexlab{a}})\citenamefont {Khayamyan}, \citenamefont
  {Lundstr{\"o}m}, \citenamefont {Gren}, \citenamefont {Lycksam},\ and\
  \citenamefont {Hellstr{\"o}m}}]{Khayamyan2017a}%
  \BibitemOpen
  \bibfield  {author} {\bibinfo {author} {\bibfnamefont {S.}~\bibnamefont
  {Khayamyan}}, \bibinfo {author} {\bibfnamefont {T.~S.}\ \bibnamefont
  {Lundstr{\"o}m}}, \bibinfo {author} {\bibfnamefont {P.}~\bibnamefont {Gren}},
  \bibinfo {author} {\bibfnamefont {H.}~\bibnamefont {Lycksam}}, \ and\
  \bibinfo {author} {\bibfnamefont {J.~G.~I.}\ \bibnamefont {Hellstr{\"o}m}},\
  }\bibfield  {title} {\enquote {\bibinfo {title} {Transitional and turbulent
  flow in a bed of spheres as measured with stereoscopic particle image
  velocimetry},}\ }\href {\doibase 10.1007/s11242-017-0819-y} {\bibfield
  {journal} {\bibinfo  {journal} {Transport in Porous Media}\ }\textbf
  {\bibinfo {volume} {117}},\ \bibinfo {pages} {45--67} (\bibinfo {year}
  {2017}{\natexlab{a}})}\BibitemShut {NoStop}%
\bibitem [{\citenamefont {Khayamyan}\ \emph
  {et~al.}(2017{\natexlab{b}})\citenamefont {Khayamyan}, \citenamefont
  {Lundstr{\"o}m}, \citenamefont {Hellstr{\"o}m}, \citenamefont {Gren},\ and\
  \citenamefont {Lycksam}}]{Khayamyan2017b}%
  \BibitemOpen
  \bibfield  {author} {\bibinfo {author} {\bibfnamefont {S.}~\bibnamefont
  {Khayamyan}}, \bibinfo {author} {\bibfnamefont {T.~S.}\ \bibnamefont
  {Lundstr{\"o}m}}, \bibinfo {author} {\bibfnamefont {J.~G.~I.}\ \bibnamefont
  {Hellstr{\"o}m}}, \bibinfo {author} {\bibfnamefont {P.}~\bibnamefont {Gren}},
  \ and\ \bibinfo {author} {\bibfnamefont {H.}~\bibnamefont {Lycksam}},\
  }\bibfield  {title} {\enquote {\bibinfo {title} {Measurements of transitional
  and turbulent flow in a randomly packed bed of spheres with particle image
  velocimetry},}\ }\href {\doibase 10.1007/s11242-016-0781-0} {\bibfield
  {journal} {\bibinfo  {journal} {Transport in Porous Media}\ }\textbf
  {\bibinfo {volume} {116}},\ \bibinfo {pages} {413--431} (\bibinfo {year}
  {2017}{\natexlab{b}})}\BibitemShut {NoStop}%
\bibitem [{\citenamefont {Larsson}, \citenamefont {Lundstr{\"o}m},\ and\
  \citenamefont {Lycksam}(2018)}]{Larsson2018}%
  \BibitemOpen
  \bibfield  {author} {\bibinfo {author} {\bibfnamefont {I.~A.~S.}\
  \bibnamefont {Larsson}}, \bibinfo {author} {\bibfnamefont {T.~S.}\
  \bibnamefont {Lundstr{\"o}m}}, \ and\ \bibinfo {author} {\bibfnamefont
  {H.}~\bibnamefont {Lycksam}},\ }\bibfield  {title} {\enquote {\bibinfo
  {title} {Tomographic piv of flow through ordered thin porous media},}\ }\href
  {\doibase 10.1007/s00348-018-2548-6} {\bibfield  {journal} {\bibinfo
  {journal} {Experiments in Fluids}\ }\textbf {\bibinfo {volume} {59}}
  (\bibinfo {year} {2018}),\ 10.1007/s00348-018-2548-6}\BibitemShut {NoStop}%
\bibitem [{\citenamefont {Gladden}\ \emph {et~al.}(2006)\citenamefont
  {Gladden}, \citenamefont {Akpa}, \citenamefont {Anadon}, \citenamefont
  {Heras}, \citenamefont {Holland}, \citenamefont {Mantle}, \citenamefont
  {Matthews}, \citenamefont {Mueller}, \citenamefont {Sains},\ and\
  \citenamefont {Sederman}}]{Gladden2006}%
  \BibitemOpen
  \bibfield  {author} {\bibinfo {author} {\bibfnamefont {L.~F.}\ \bibnamefont
  {Gladden}}, \bibinfo {author} {\bibfnamefont {B.~S.}\ \bibnamefont {Akpa}},
  \bibinfo {author} {\bibfnamefont {L.~D.}\ \bibnamefont {Anadon}}, \bibinfo
  {author} {\bibfnamefont {J.~J.}\ \bibnamefont {Heras}}, \bibinfo {author}
  {\bibfnamefont {D.~J.}\ \bibnamefont {Holland}}, \bibinfo {author}
  {\bibfnamefont {M.~D.}\ \bibnamefont {Mantle}}, \bibinfo {author}
  {\bibfnamefont {S.}~\bibnamefont {Matthews}}, \bibinfo {author}
  {\bibfnamefont {C.}~\bibnamefont {Mueller}}, \bibinfo {author} {\bibfnamefont
  {M.~C.}\ \bibnamefont {Sains}}, \ and\ \bibinfo {author} {\bibfnamefont
  {A.~J.}\ \bibnamefont {Sederman}},\ }\bibfield  {title} {\enquote {\bibinfo
  {title} {Dynamic mr imaging of single- and two-phase flows},}\ }\href
  {\doibase 10.1205/cherd06019} {\bibfield  {journal} {\bibinfo  {journal}
  {Chemical Engineering Research and Design}\ }\textbf {\bibinfo {volume}
  {84}},\ \bibinfo {pages} {272--281} (\bibinfo {year} {2006})}\BibitemShut
  {NoStop}%
\bibitem [{\citenamefont {Poelma}(2020)}]{Poelma2020}%
  \BibitemOpen
  \bibfield  {author} {\bibinfo {author} {\bibfnamefont {C.}~\bibnamefont
  {Poelma}},\ }\bibfield  {title} {\enquote {\bibinfo {title} {Measurement in
  opaque flows: a review of measurement techniques for dispersed multiphase
  flows},}\ }\href {\doibase 10.1007/s00707-020-02683-x.} {\bibfield  {journal}
  {\bibinfo  {journal} {Acta mechanica}\ }\textbf {\bibinfo {volume} {231}},\
  \bibinfo {pages} {2089--2111} (\bibinfo {year} {2020})}\BibitemShut {NoStop}%
\bibitem [{\citenamefont {Hassan}\ and\ \citenamefont
  {Dominguez-Ontiveros}(2008)}]{Hassan2008}%
  \BibitemOpen
  \bibfield  {author} {\bibinfo {author} {\bibfnamefont {Y.~A.}\ \bibnamefont
  {Hassan}}\ and\ \bibinfo {author} {\bibfnamefont {E.~E.}\ \bibnamefont
  {Dominguez-Ontiveros}},\ }\bibfield  {title} {\enquote {\bibinfo {title}
  {Flow visualization in a pebble bed reactor experiment using piv and
  refractive index matching techniques},}\ }\href {\doibase
  10.1016/j.nucengdes.2008.01.027} {\bibfield  {journal} {\bibinfo  {journal}
  {Nuclear Engineering and Design}\ }\textbf {\bibinfo {volume} {238}},\
  \bibinfo {pages} {3080--3085} (\bibinfo {year} {2008})}\BibitemShut {NoStop}%
\bibitem [{\citenamefont {Kov{\'a}ts}, \citenamefont {Th{\'e}venin},\ and\
  \citenamefont {Z{\"a}hringer}(2015)}]{Kovats2015}%
  \BibitemOpen
  \bibfield  {author} {\bibinfo {author} {\bibfnamefont {P.}~\bibnamefont
  {Kov{\'a}ts}}, \bibinfo {author} {\bibfnamefont {D.}~\bibnamefont
  {Th{\'e}venin}}, \ and\ \bibinfo {author} {\bibfnamefont {K.}~\bibnamefont
  {Z{\"a}hringer}},\ }\bibfield  {title} {\enquote {\bibinfo {title}
  {Experimental investigation of flow fields within cavities of coarse packed
  bed},}\ }\href@noop {} {\bibfield  {journal} {\bibinfo  {journal} {Fachtagung
  {\textquotedbl}Lasermethoden in der Str{\"o}mungsmesstechnik{\textquotedbl}}\
  ,\ \bibinfo {pages} {54--1--7}} (\bibinfo {year} {8.-10. September
  2015})}\BibitemShut {NoStop}%
\bibitem [{\citenamefont {Martins}\ \emph {et~al.}(2018)\citenamefont
  {Martins}, \citenamefont {{Da Carvalho Silva}}, \citenamefont {Lessig},\ and\
  \citenamefont {Z{\"a}hringer}}]{Martins2018}%
  \BibitemOpen
  \bibfield  {author} {\bibinfo {author} {\bibfnamefont {F.}~\bibnamefont
  {Martins}}, \bibinfo {author} {\bibfnamefont {C.}~\bibnamefont {{Da Carvalho
  Silva}}}, \bibinfo {author} {\bibfnamefont {C.}~\bibnamefont {Lessig}}, \
  and\ \bibinfo {author} {\bibfnamefont {K.}~\bibnamefont {Z{\"a}hringer}},\
  }\bibfield  {title} {\enquote {\bibinfo {title} {Ray-tracing based image
  correction of optical distortion for piv measurements in packed beds},}\
  }\href {\doibase 10.32604/jaop.2018.03870} {\bibfield  {journal} {\bibinfo
  {journal} {Journal of Advanced Optics and Photonics}\ }\textbf {\bibinfo
  {volume} {1}},\ \bibinfo {pages} {71--94} (\bibinfo {year}
  {2018})}\BibitemShut {NoStop}%
\bibitem [{\citenamefont {Velten}\ \emph {et~al.}(2024)\citenamefont {Velten},
  \citenamefont {Ebert}, \citenamefont {Lessig},\ and\ \citenamefont
  {{Katharina Z{\"a}hringer}}}]{Velten2024}%
  \BibitemOpen
  \bibfield  {author} {\bibinfo {author} {\bibfnamefont {C.}~\bibnamefont
  {Velten}}, \bibinfo {author} {\bibfnamefont {M.}~\bibnamefont {Ebert}},
  \bibinfo {author} {\bibfnamefont {C.}~\bibnamefont {Lessig}}, \ and\ \bibinfo
  {author} {\bibnamefont {{Katharina Z{\"a}hringer}}},\ }\bibfield  {title}
  {\enquote {\bibinfo {title} {Ray tracing particle image velocimetry --
  challenges in the application to a packed bed},}\ }\href {\doibase
  10.1016/j.partic.2023.06.003} {\bibfield  {journal} {\bibinfo  {journal}
  {Particuology}\ }\textbf {\bibinfo {volume} {84}},\ \bibinfo {pages}
  {194--208} (\bibinfo {year} {2024})}\BibitemShut {NoStop}%
\bibitem [{\citenamefont {Whitaker}(1999)}]{whitaker1999}%
  \BibitemOpen
  \bibfield  {author} {\bibinfo {author} {\bibfnamefont {S.}~\bibnamefont
  {Whitaker}},\ }\href {\doibase 10.1007/978-94-017-3389-2} {\emph {\bibinfo
  {title} {The {{Method}} of {{Volume Averaging}}}}},\ edited by\ \bibinfo
  {editor} {\bibfnamefont {J.}~\bibnamefont {Bear}},\ \bibinfo {series} {Theory
  and {{Applications}} of {{Transport}} in {{Porous Media}}}, Vol.~\bibinfo
  {volume} {13}\ (\bibinfo  {publisher} {{Springer Netherlands}},\ \bibinfo
  {address} {{Dordrecht}},\ \bibinfo {year} {1999})\BibitemShut {NoStop}%
\bibitem [{\citenamefont {Jasak}(1996)}]{jasak1996}%
  \BibitemOpen
  \bibfield  {author} {\bibinfo {author} {\bibfnamefont {H.}~\bibnamefont
  {Jasak}},\ }\emph {\bibinfo {title} {Error {{Analysis}} and {{Estimation}}
  for the {{Finite Volume Method}} with {{Applications}} to {{Fluid Flows}}}},\
  \href@noop {} {\bibinfo {type} {Doctoral}},\ \bibinfo  {school} {Imperial
  College}, \bibinfo {address} {{London}} (\bibinfo {year} {1996})\BibitemShut
  {NoStop}%
\bibitem [{\citenamefont {Moukalled}, \citenamefont {Mangani},\ and\
  \citenamefont {Darwish}(2016)}]{moukalled2016}%
  \BibitemOpen
  \bibfield  {author} {\bibinfo {author} {\bibfnamefont {F.}~\bibnamefont
  {Moukalled}}, \bibinfo {author} {\bibfnamefont {L.}~\bibnamefont {Mangani}},
  \ and\ \bibinfo {author} {\bibfnamefont {M.}~\bibnamefont {Darwish}},\ }\href
  {\doibase 10.1007/978-3-319-16874-6} {\emph {\bibinfo {title} {The {{Finite
  Volume Method}} in {{Computational Fluid Dynamics}}: {{An Advanced
  Introduction}} with {{OpenFOAM}}\textregistered{} and {{Matlab}}}}},\
  \bibinfo {series} {Fluid {{Mechanics}} and {{Its Applications}}}, Vol.\
  \bibinfo {volume} {113}\ (\bibinfo  {publisher} {{Springer International
  Publishing}},\ \bibinfo {address} {{Cham}},\ \bibinfo {year}
  {2016})\BibitemShut {NoStop}%
\bibitem [{\citenamefont {Komen}\ \emph {et~al.}(2017)\citenamefont {Komen},
  \citenamefont {Camilo}, \citenamefont {Shams}, \citenamefont {Geurts},\ and\
  \citenamefont {Koren}}]{komen2017}%
  \BibitemOpen
  \bibfield  {author} {\bibinfo {author} {\bibfnamefont {E.}~\bibnamefont
  {Komen}}, \bibinfo {author} {\bibfnamefont {L.}~\bibnamefont {Camilo}},
  \bibinfo {author} {\bibfnamefont {A.}~\bibnamefont {Shams}}, \bibinfo
  {author} {\bibfnamefont {B.}~\bibnamefont {Geurts}}, \ and\ \bibinfo {author}
  {\bibfnamefont {B.}~\bibnamefont {Koren}},\ }\bibfield  {title} {\enquote
  {\bibinfo {title} {A quantification method for numerical dissipation in
  quasi-{DNS} and under-resolved {DNS}, and effects of numerical dissipation in
  quasi-{DNS} and under-resolved {DNS} of turbulent channel flows},}\ }\href
  {\doibase 10.1016/j.jcp.2017.05.030} {\bibfield  {journal} {\bibinfo
  {journal} {Journal of Computational Physics}\ }\textbf {\bibinfo {volume}
  {345}},\ \bibinfo {pages} {565--595} (\bibinfo {year} {2017})}\BibitemShut
  {NoStop}%
\bibitem [{\citenamefont {{ANSYS, Inc.}}(2023)}]{ansysfl}%
  \BibitemOpen
  \bibfield  {author} {\bibinfo {author} {\bibnamefont {{ANSYS, Inc.}}},\
  }\href@noop {} {\emph {\bibinfo {title} {ANSYS Fluent User's Guide, Release
  2023 R2}}} (\bibinfo {year} {2023})\BibitemShut {NoStop}%
\bibitem [{\citenamefont {Komen}\ \emph {et~al.}(2014)\citenamefont {Komen},
  \citenamefont {Shams}, \citenamefont {Camilo},\ and\ \citenamefont
  {Koren}}]{komen2014}%
  \BibitemOpen
  \bibfield  {author} {\bibinfo {author} {\bibfnamefont {E.}~\bibnamefont
  {Komen}}, \bibinfo {author} {\bibfnamefont {A.}~\bibnamefont {Shams}},
  \bibinfo {author} {\bibfnamefont {L.}~\bibnamefont {Camilo}}, \ and\ \bibinfo
  {author} {\bibfnamefont {B.}~\bibnamefont {Koren}},\ }\bibfield  {title}
  {\enquote {\bibinfo {title} {Quasi-{{DNS}} capabilities of {{OpenFOAM}} for
  different mesh types},}\ }\href {\doibase 10.1016/j.compfluid.2014.02.013}
  {\bibfield  {journal} {\bibinfo  {journal} {Computers \& Fluids}\ }\textbf
  {\bibinfo {volume} {96}},\ \bibinfo {pages} {87--104} (\bibinfo {year}
  {2014})}\BibitemShut {NoStop}%
\bibitem [{\citenamefont {Issa}(1986)}]{issa1986}%
  \BibitemOpen
  \bibfield  {author} {\bibinfo {author} {\bibfnamefont {R.~I.}\ \bibnamefont
  {Issa}},\ }\bibfield  {title} {\enquote {\bibinfo {title} {Solution of the
  implicitly discretised fluid flow equations by operator-splitting},}\ }\href
  {\doibase https://doi.org/10.1016/0021-9991(86)90099-9} {\bibfield  {journal}
  {\bibinfo  {journal} {Journal of Computational Physics}\ }\textbf {\bibinfo
  {volume} {62}},\ \bibinfo {pages} {40--65} (\bibinfo {year}
  {1986})}\BibitemShut {NoStop}%
\bibitem [{\citenamefont {{LaVision GmbH}}(2019)}]{lavision2019}%
  \BibitemOpen
  \bibfield  {author} {\bibinfo {author} {\bibnamefont {{LaVision GmbH}}},\
  }\href {https://www.lavision.de/de/download.php?id=3020} {\enquote {\bibinfo
  {title} {8.4 davis software manual},}\ } (\bibinfo {year}
  {Jul-2019})\BibitemShut {NoStop}%
\bibitem [{\citenamefont {Wilcox}(2006)}]{wilcox2006}%
  \BibitemOpen
  \bibfield  {author} {\bibinfo {author} {\bibfnamefont {D.~C.}\ \bibnamefont
  {Wilcox}},\ }\href@noop {} {\emph {\bibinfo {title} {Turbulence modeling for
  {CFD}}}},\ \bibinfo {edition} {3rd}\ ed.\ (\bibinfo  {publisher} {DCW
  Industries},\ \bibinfo {address} {La C\~{a}nada, Calif},\ \bibinfo {year}
  {2006})\BibitemShut {NoStop}%
\end{thebibliography}%

\end{document}